\documentclass[superscriptaddress, twocolumn,showpacs,
amssymb,amsmath,nobibnotes,aps,prd,
nofootinbib]{revtex4-1}
\pdfoutput=1
\usepackage{graphicx,subfigure,bm,color,psfrag,hyperref}
\usepackage{amsfonts}
\usepackage{lipsum}
\usepackage{mathtools}
\usepackage{verbatim}
\usepackage[normalem]{ulem}
\usepackage[dvipsnames]{xcolor}
\hypersetup{colorlinks,linkcolor={blue},citecolor={red},urlcolor={magenta}}   

\begin{document}


\title{Late-Time Emergence of Dark Energy and Its Interaction with Dark Matter}


\author{Sibo Zhang}
\email{sbzhang02@163.com}
\affiliation{Department of Physics, Liaoning Normal University, Dalian, 116029, People's Republic of China}

\author{Weiqiang Yang}
\email{d11102004@163.com}
\affiliation{Department of Physics, Liaoning Normal University, Dalian, 116029, People's Republic of China}

\author{Supriya Pan}
\email{supriya.maths@presiuniv.ac.in}
\affiliation{Department of Mathematics, Presidency University, 86/1 College Street, Kolkata 700073, India}
\affiliation{Institute of Systems Science, Durban University of Technology, Durban 4000, Republic of South Africa}

\author{Olga Mena}
\email{omena@ific.uv.es}
\affiliation{Instituto de F\'{i}sica Corpuscular (CSIC-Universitat de Val\`{e}ncia), E-46980 Paterna, Spain} 

\author{Eleonora Di Valentino}
\email{e.divalentino@sheffield.ac.uk}
\affiliation{School of Mathematical and Physical Sciences, University of Sheffield, Hounsfield Road, Sheffield S3 7RH, United Kingdom}

\author{Subenoy Chakraborty}
\email{schakraborty.math@gmail.com}
\affiliation{Department of Mathematics, Brainware University, Barasat, West Bengal 700125, India}

\begin{abstract}
 We present an interacting scenario between dark energy (DE) and dark matter (DM), where DE has an emergent nature, that means, DE was absent in the early universe but it becomes effective only at late times. We consider two specific emergent DE models, one with no free parameters and the other featuring two parameters describing the speed and epoch of the transition.
We constrain  both scenarios using the cosmic microwave background (CMB) measurements from the Planck 2018 release, baryon acoustic oscillations from DESI DR2, and three different compilations of Type Ia supernovae (PantheonPlus, DES-Dovekie, and Union3). Our analysis indicates that current cosmological probes are not able to tightly constrain the speed of the transition.
For both scenarios, the posterior distribution of the interaction parameter is shifted away from zero at more than 95\% CL whenever the CMB data are combined with any of these additional probes, with the preferred direction corresponding to a transfer of energy from DE to DM.
While CMB alone yields a high value of $H_0$, in agreement with local determinations, this effect is reduced when DESI is added and disappears once supernova data are included. In contrast, the clustering parameter $S_8$ is consistently shifted toward lower values in the combined datasets, and it is correlated with the preference for a negative interaction.  However, according to the $\Delta \chi^2_{\rm min}$ and Bayesian evidence, none of the interacting models is favored over $\Lambda$CDM or $w_0w_a$CDM, indicating that the interaction does not rescue these emergent DE models.  Our results therefore highlight the limitations of these scenarios in addressing current cosmological tensions, while pointing to the crucial role of future data in further assessing their viability. 
 
\end{abstract}

\maketitle
\section{Introduction}

According to the observational evidence available so far, our Universe is currently expanding with acceleration~\cite{SupernovaSearchTeam:1998fmf,SupernovaCosmologyProject:1998vns}. This accelerating phase is driven either by some hypothetical dark energy (DE)~\cite{Copeland:2006wr,Bamba:2012cp} or by a geometrical DE~\cite{Nojiri:2006ri,DeFelice:2010aj,Capozziello:2011et,Clifton:2011jh,Cai:2015emx,Nojiri:2017ncd,Bahamonde:2021gfp} component, but the origin of this dark component is not yet clearly understood. Observational data further suggest that nearly 96\% of the total energy budget of our Universe consists of dark matter (DM) and DE (or geometrical DE), with about 68\% attributed to DE (or geometrical DE). DM is responsible for the observed structure formation of the Universe. 
This information is well described by the $\Lambda$ Cold Dark Matter ($\Lambda$CDM) cosmological model, in which the cosmological constant $\Lambda$, inserted into Einstein's General Relativity, acts as the source of DE, while DM is pressureless. In the $\Lambda$CDM framework, DM and DE are conserved separately; that is, apart from gravitational interaction, these components do not interact with each other. According to observational evidence, the $\Lambda$CDM model has been quite successful in explaining a large number of astronomical surveys. However, based on past and present cosmological records, $\Lambda$CDM faces several theoretical and observational challenges. It is strongly argued that $\Lambda$CDM may be only an approximation of a more realistic theory capable of explaining all observational discrepancies~\cite{Abdalla:2022yfr,CosmoVerseNetwork:2025alb}.

One of the natural generalizations of the $\Lambda$CDM model is to consider a mutual interaction between DM and DE. The possibility of such an interaction has been widely investigated in the literature~\cite{Amendola:1999er,Barrow:2006hia,Valiviita:2009nu,Clemson:2011an,Yang:2014gza,
DiValentino:2017iww,DiValentino:2019ffd,Yang:2021hxg,Giare:2024smz,Silva:2025hxw,Pan:2025qwy,Paliathanasis:2026ymi} (see the reviews on interacting DE~\cite{Bolotin:2013jpa,Wang:2016lxa,Wang:2024vmw}).
In particular, it has been found that interacting cosmologies may weaken the cosmic coincidence problem~\cite{Amendola:1999er,Cai:2004dk,Pavon:2005yx,Huey:2004qv,delCampo:2008sr,delCampo:2008jx}, allow the crossing of the phantom divide line without invoking scalar field theory~\cite{Wang:2005jx,Sadjadi:2006qb,Pan:2014afa}, alleviate the Hubble constant tension---one of the greatest puzzles of the present time~\cite{Kumar:2017dnp,DiValentino:2017iww,Yang:2018euj,Pan:2019gop,Giare:2024smz}---and reduce the clustering tension~\cite{Pourtsidou:2016ico,An:2017crg,Kumar:2019wfs}. Thus, the interaction framework can be considered a strong generalization of the non-interacting $\Lambda$CDM model. 
In interacting cosmologies, the DE equation of state is usually taken to be either constant~\cite{Pan:2016ngu,Yang:2018euj,Pan:2020bur,Yang:2021oxc} or time-dependent~\cite{Valiviita:2009nu,Pan:2016ngu,Yang:2018uae,Pan:2019gop}. In this work, we investigate an interacting scenario between DM and DE in which DE has an emergent nature in recent times; that is, DE had no effective presence in the early Universe but emerges at late times. As the DE equation of state is dynamical, therefore the underlying interacting scenario generalizes those with a constant DE equation of state.  
The emergence of DE has recently received significant attention in the community~\cite{Li:2019yem,Pan:2019hac,Li:2020ybr,Yang:2020ope,Rezaei:2020mrj,Hernandez-Almada:2020uyr,Benaoum:2020qsi,Yang:2021eud}, but the possibility of an interaction between emergent DE and DM has never been reported in the literature. We consider two emergent DE models: one with no free parameters~\cite{Li:2019yem,Pan:2019hac}, and another, newly introduced in this work, featuring two free parameters corresponding to the epoch and the speed of the transition. 
 
The article is organized as follows. In Section~\ref{sec-2}, we introduce the emergent DE and the interaction scenario. In Section~\ref{sec-data}, we describe the observational datasets and the fitting technique used to constrain the proposed scenario. Section~\ref{sec-results} presents the results of the interacting scenario. Finally, Section~\ref{sec-conclusion} provides an overall summary and conclusions.

\section{Interacting Emergent dark energy}
\label{sec-2}

We consider that the geometry of our Universe on large scales is described by the spatially flat Friedmann--Lema\^{i}tre--Robertson--Walker (FLRW) line element 
\begin{equation}
ds^2 = -dt^2 + a^2(t) (dx^2 + dy^2 + dz^2),
\end{equation}
where $(t, x, y, z)$ are the co-moving coordinates, and $a(t)$ is the expansion scale factor of the Universe, which depends on the cosmic time $t$. We further assume that the gravitational sector of the Universe is described by Einstein's General Relativity (GR), while the matter sector is composed of baryons, radiation, pressureless DM, and DE. Among these four components, only DM and DE are allowed to interact with each other. 

The conservation equations of the interacting DM--DE fluids take the forms  
\begin{eqnarray}
&&\frac{d\rho_{\rm DM}}{dt} = - 3 H \rho_{\rm DM} -Q,\label{cons1}\\
&&\frac{d\rho_{\rm DE}}{dt} = - 3 H (1+w_{\rm  DE}) \rho_{\rm DE} + Q,\label{cons2}
\end{eqnarray} 
where $w_{\rm DE}$ is the barotropic equation-of-state (EoS) parameter of the DE fluid, $Q$ describes the interaction rate between the two sectors, and $H$ is the Hubble parameter of the FLRW Universe, which is constrained by the Friedmann equation
\begin{eqnarray}\label{Hubble}
H^2  = \frac{1}{3M_p^2} \sum_{i} \rho_i,
\end{eqnarray}
with $M_{p}^2 = (8 \pi G)^{-1}$ being the reduced Planck mass. Here, $\rho$ denotes the energy density of a fluid, and $\rho_i$ specifically represents the energy density of the $i$-th component. We assume that $w_{\rm DE}$ is time-dependent. Note that for $Q>0$, energy is transferred from DM to DE, while $Q<0$ indicates energy transfer in the opposite direction, i.e.\ from DE to DM.
Once the interaction function $Q$ and the DE equation of state $w_{\rm DE}$ are specified, the expansion history of the Universe can be determined either analytically or numerically, depending on the form of $Q$. Concerning the choice of the interaction function, one may in principle adopt an arbitrary form, since no fundamental action formalism is available yet. However, in the present work we consider the most well-known interaction function,  
\begin{eqnarray}\label{interaction-function}
Q = 3 H \xi \rho_{\rm DE},
\end{eqnarray} 
where $\xi$ is the dimensionless coupling parameter of the interaction. According to the rule of energy transfer, $\xi > 0$ corresponds to the flow of energy from DM to DE, while $\xi < 0$ denotes the opposite flow, from DE to DM.  
It is worth noting that the above choice is not made purely on phenomenological grounds. In fact, this form of interaction can be recovered from scalar field potentials~\cite{Pan:2020zza}. Moreover, it has been explicitly shown in~\cite{Pan:2020mst} that such interaction functions can also arise in other cosmological theories. We also refer to a series of earlier works, where it has been demonstrated that interaction models can be derived from an action principle~\cite{vandeBruck:2015ida,Boehmer:2015kta,Boehmer:2015sha,Gleyzes:2015pma}.

It is interesting to note that any coupled DE--DM system can be recast as an equivalent uncoupled system  of DE and DM. This can be seen by rewriting the conservation equations (\ref{cons1}) and (\ref{cons2}) in terms of the effective EoS of the dark fluids:  

\begin{align}
\frac{d\rho_{\rm DM}}{dt} &= - 3 H \rho_{\rm DM} \left(1 + w_{\rm DM}^{\rm eff} \right), 
&\quad w_{\rm DM}^{\rm eff} = \frac{Q}{3H \rho_{\rm DM}}, \label{cons1A}
\end{align}
\begin{align}
\frac{d\rho_{\rm DE}}{dt} &= - 3 H \rho_{\rm DE} \left(1+w_{\rm DE}^{\rm eff}\right), 
&\quad w_{\rm DE}^{\rm eff} = w_{\rm DE} - \frac{Q}{3H \rho_{\rm DE}}. \label{cons2A}
\end{align}
The above equations show that an interacting scenario between DE and DM is equivalent to a non-interacting scenario in which the EoS parameters of the dark fluids become dynamical~\cite{Pan:2025qwy,Petri:2025swg}. On the DM side, $w_{\rm DM}^{\rm eff}$ can take either positive or negative values depending on the sign of the coupling function.  
On the DE side, the effective EoS is also modified. In particular, if $w_{\rm DE} > -1$, one can obtain an effective DE scenario with $w_{\rm DE}^{\rm eff} < -1$, or it may remain in the quintessence regime (if the interaction is not strong enough to drive $w_{\rm DE}^{\rm eff}$ into the phantom regime), or the effective scenario can mimic $w_{\rm DE}^{\rm eff} = -1$.\footnote{One can easily verify that an effective description with $w_{\rm DE}^{\rm eff} = -1$ can be obtained with a suitable choice of the interaction, $Q = 3 H (1+w_{\rm DE}) \rho_{\rm DE}$. This indicates that the coupling parameter of the interaction should depend on the EoS of DE.} Similar situations can arise for the cases $w_{\rm DE} < -1$ or $w_{\rm DE} = -1$. These behaviours depend on both the strength and the sign of the interaction.

At this point, we would like to remark that the equivalence established through Eqs.~(\ref{cons1A}) and (\ref{cons2A}) implies an additional insight. 
Without introducing a dynamical EoS for DE, i.e.\ by considering a constant DE EoS, one can obtain an identical effective EoS for DE through the introduction of a dynamical coupling, $\xi(a)$. Interacting models with a dynamical coupling have been investigated recently, where the DE EoS has been taken either as $-1$~\cite{Yang:2019uzo,Yang:2025uyv} or as a constant different from $-1$~\cite{Yang:2020tax}. 
In particular, in Ref.~\cite{Yang:2025uyv}, assuming $w_{\rm DE} = -1$ as the simplest case, the interacting model $Q = 3 H \xi(a) \rho_{\rm DE}$ has been confronted with recent observational data for two different choices of the coupling function: one in which $\xi(a)$ takes a linear form in the scale factor, $\xi(a) = \xi_0 + \xi_a (1-a)$ (with $\xi_0$ and $\xi_a$ constants), and another with a single free parameter given by $\xi(a) = \xi_0 \left(1 + \frac{1-a}{a^2 + (1-a)^2} \right)$~\cite{Yang:2025uyv}. 
Additionally, we note that, with a dynamical coupling $\xi(a)$, the effective EoS of DM also depends on the interaction function $Q$, and therefore its evolution is determined by the specific choice of the coupling function.

To understand the full picture of the interacting model, one needs to investigate its behaviour in the large scale of the universe. Thus, we now move to the investigation of the interacting scenario at the level of perturbations. 
We consider the perturbed line element in the synchronous gauge: $ds^2 = a^2(\tau) \left [-d\tau^2 + (\delta_{ij}+h_{ij}) dx^i dx^j  \right]$, 
where $\tau$ is the conformal time, $\delta_{ij}$ ($h_{ij}$) corresponds to the unperturbed (perturbed) metric tensor. For this metric, one can write the evolution equations for the density perturbations and the velocity divergences in the Fourier space as follows: 
\begin{widetext}
	\begin{eqnarray}
		&&\delta _{\rm DE}^{\prime } =-(1+w_{\rm DE})\left( \theta _{\rm DE}+\frac{h^{\prime }}{2}%
		\right) -3\mathcal{H}(c_{s, {\rm DE}}^{2}-w_{\rm DE})\left[ \delta _{\rm DE}+3\mathcal{H}%
		(1+w_{\rm DE})\frac{\theta _{\rm DE}}{k^{2}}\right] \\
		&&~~~~~-3\mathcal{H}w_{\rm DE}^{\prime}\frac{\theta_{\rm DE}}{k^{2}}
		+9\mathcal{H}^2\xi(c_{s,{\rm DE}}^{2}-w_{\rm DE})\frac{\theta _{\rm DE}}{k^{2}}, \\
		&&\theta _{\rm DE}^{\prime } =-\mathcal{H}(1-3c_{s,{\rm DE}}^{2})\theta _{\rm DE}+\frac{%
			c_{s,{\rm DE}}^{2}}{(1+w_{\rm DE})}k^{2}\delta _{\rm DE}+3\mathcal{H}\xi\left[ \frac{%
			\theta _{\rm DM}-(1+c_{s,{\rm DE}}^{2})\theta _{\rm DE}}{1+w_{\rm DE}}\right], \\
		&&\delta _{\rm DM}^{\prime } =-\left( \theta _{\rm DM}+\frac{h^{\prime }}{2}%
		\right)+3\mathcal{H}\xi\frac{\rho _{\rm DE}}{\rho _c}\left( \delta _{\rm DM}-\delta_{\rm DE}\right), \\
		&&\theta _{\rm DM}^{\prime } =-\mathcal{H}\theta_{\rm DM},
	\end{eqnarray}
\end{widetext}
where a prime denotes the derivative with respect to $\tau$, $h$ (only here) is the trace of the metric perturbations, $k$ is the wavenumber in the Fourier
space and $c^2_{s, {\rm DE}}$ is the square of the sound speed of DE in its rest frame which we set to be unity, as usually considered in the context of minimally coupled scalar field models.

\begin{figure*}
    \centering
    \includegraphics[width=0.48\textwidth]{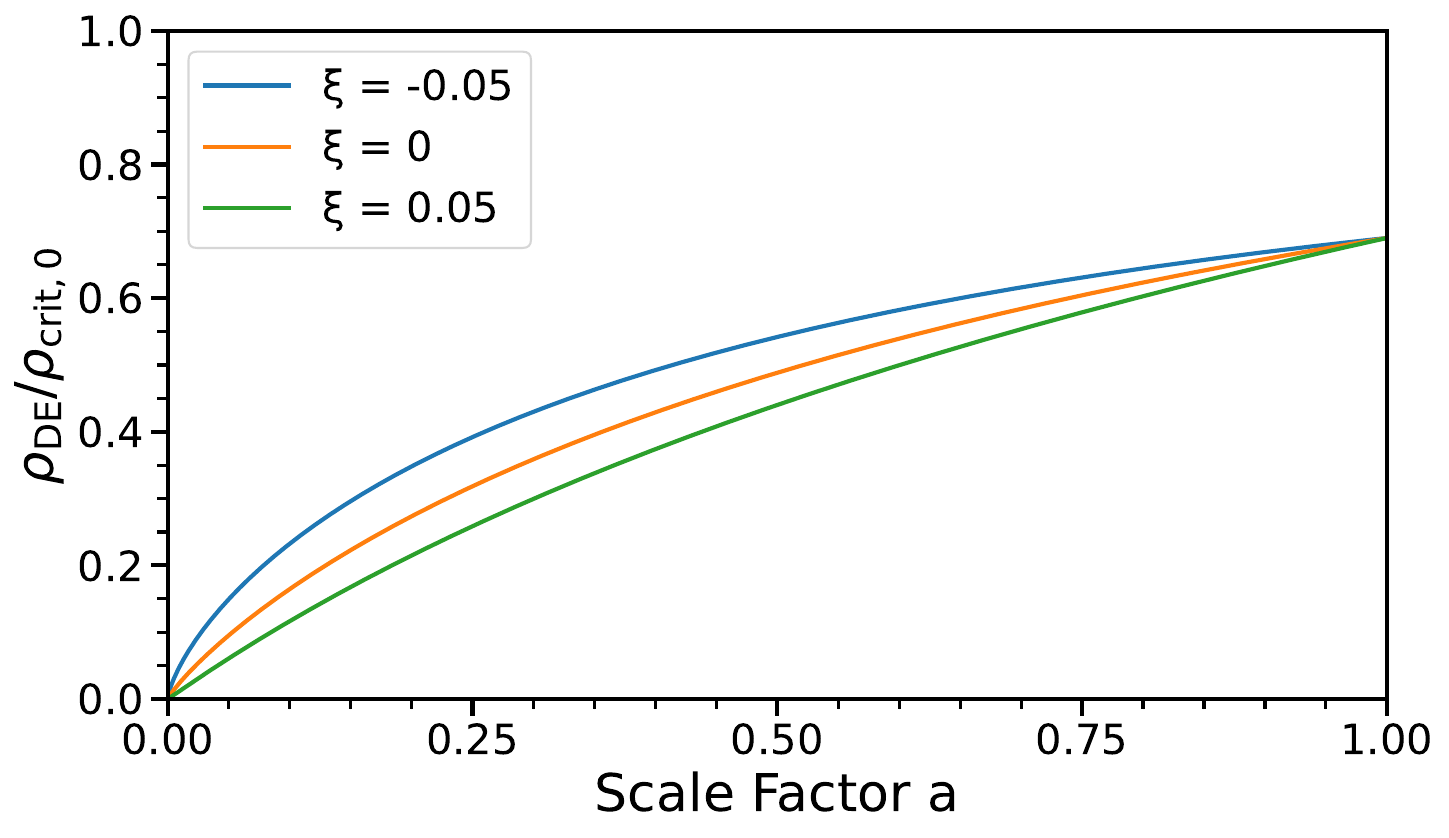}
    \includegraphics[width=0.48\textwidth]{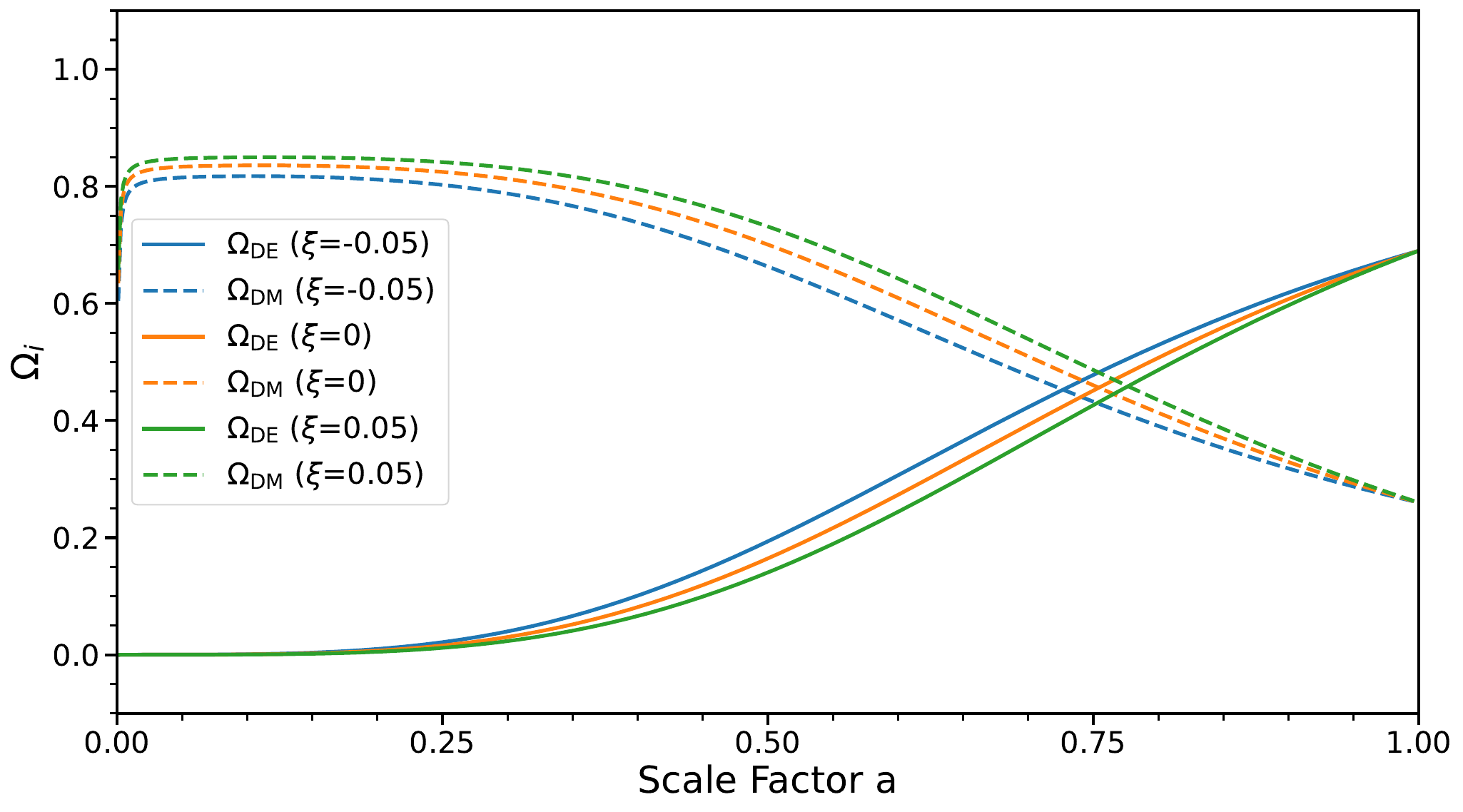}
    \caption{Evolution of $\rho_{\rm DE}/\rho_{\rm crit,0}$ (left panel) and evolution of the DE and CDM density parameters, $\Omega_{\rm DE}$ and $\Omega_{\rm CDM}$ (right panel) for the {\bf IPEDE} model. }
    \label{fig:energy-density}
\end{figure*}

\begin{figure*}
    \centering
    \includegraphics[width=0.48\textwidth]{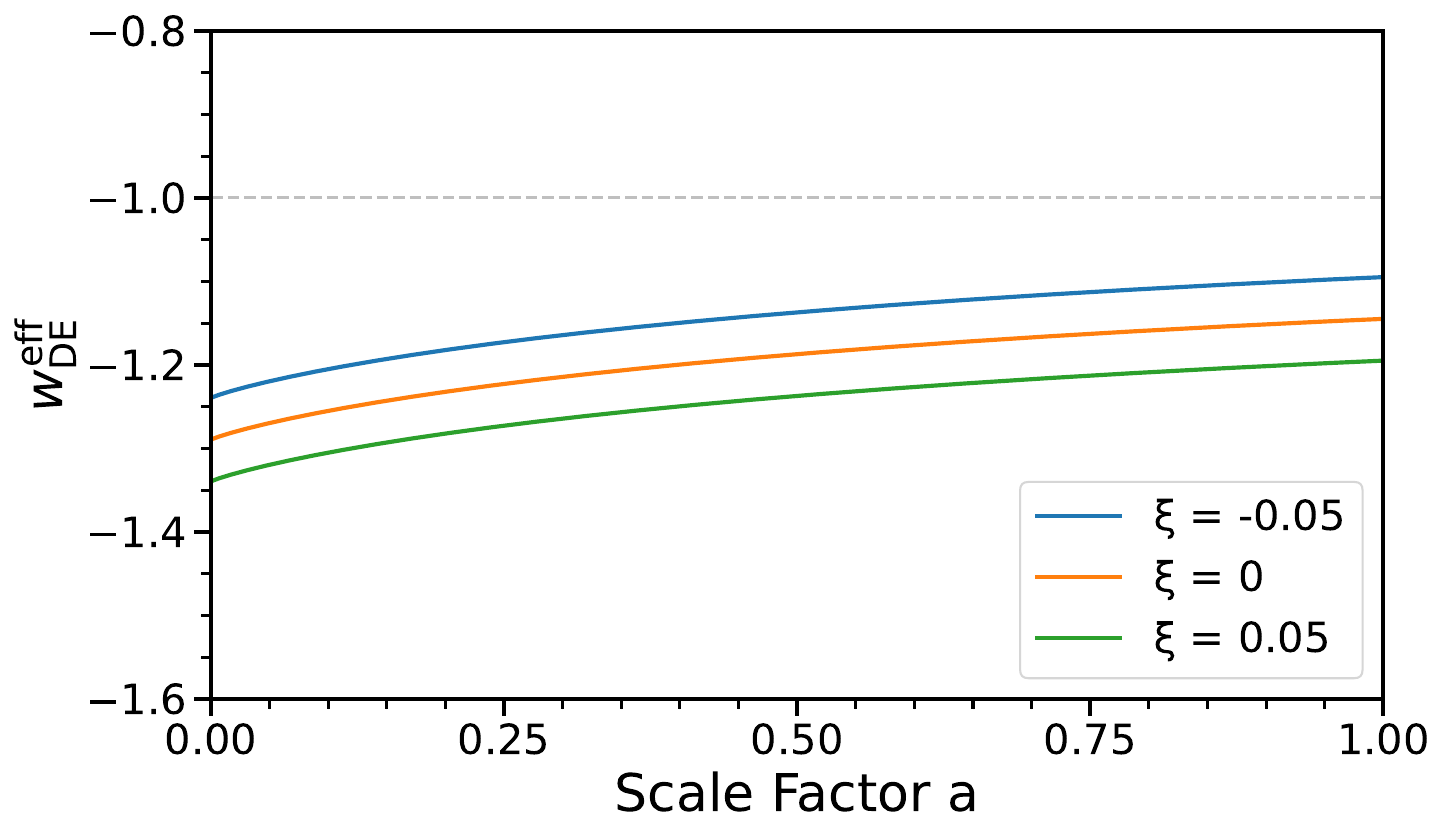}
    \includegraphics[width=0.48\textwidth]{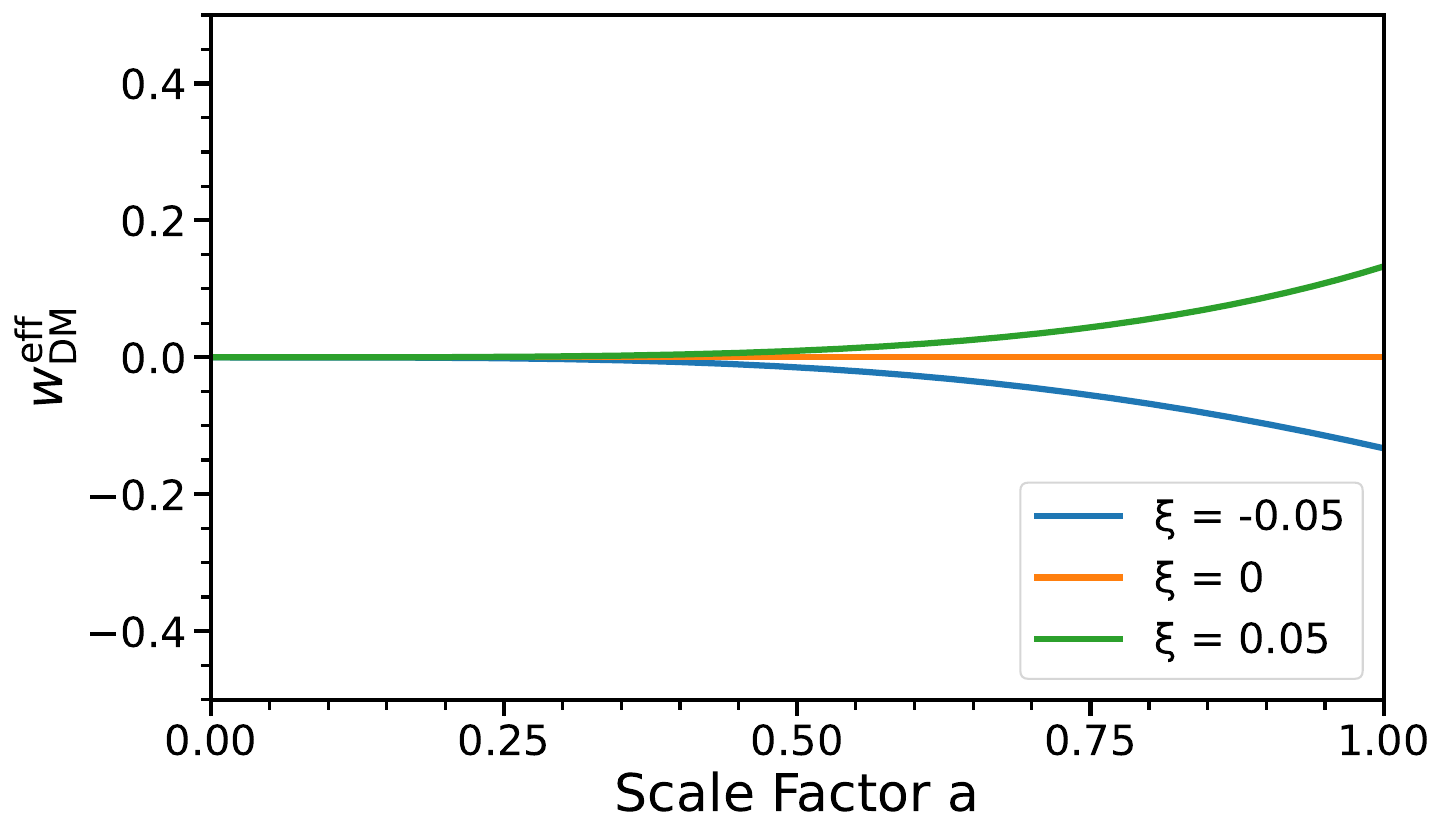}
    \caption{Evolution of the effective EoS parameters for different values of the coupling parameter $\xi$ for the {\bf IPEDE} model. The left panel shows $w_{\rm DE}^{\rm eff}$, while the right panel shows $w_{\rm DM}^{\rm eff}$. }
    \label{fig:eff-eos}
\end{figure*}

\subsection{Models}

Focusing now on the DE EoS, we consider it to be time-varying, but we do not adopt any of the usual parameterizations. A particular model of DE was recently proposed in Ref.~\cite{Li:2019yem} (see also~\cite{Pan:2019hac}, where its perturbative behavior was investigated), referred to as the phenomenologically emergent DE. In this model, DE has no effective presence in the early Universe but emerges only at recent times. This is a phenomenologically appealing proposal, since cosmic acceleration, as indicated by Type Ia supernovae and other probes, began only recently, with DE becoming dominant in the energy budget of the Universe. Along this line, a DE model that is only recently active in the cosmic setup is particularly appealing. The EoS of such an emergent DE has the following form~\cite{Li:2019yem,Pan:2019hac}:  

\begin{align}\label{eos}
w_{\rm DE}(a) = -1 - \frac{1}{3 \ln 10} \Bigl[ 1 - \tanh \bigl(\log_{10} a \bigr) \Bigr].
\end{align}
Note that the EoS in (\ref{eos}) does not contain any free parameter. Within this parametrized EoS, in the very early-epoch, i.e. $a \rightarrow 0$, $w_{\rm DE}(a) \sim -1.29$; in the present epoch, $w_{\rm DE}$ remains in the phantom regime ($w_{\rm DE} (a) = -1 - 1/(3 \ln 10)$) and in the far future, DE EoS asymptotically reaches the cosmological constant, i.e. $w_{\rm DE} (a) \sim -1$. This is not surprising since the above parameterization is purely motivated from the phenomenological ground, and thus, the values of $w_{\rm DE}$ adopted in different phases of the universe are solely dependent on the functional form of $w_{\rm DE} (a)$. While a different phenomenological parameterization of emergent DE may yield a different early-time value of $w_{\rm DE}(a)$, and therefore, without fixing the early-time value of the DE EoS, one can introduce a new emergent DE model in which the EoS of DE at the early epoch could be a free parameter.   
The resulting interacting scenario driven by (\ref{eos}) is labeled as {\bf IPEDE}.
In the left panel of Fig.~\ref{fig:energy-density}, we show the evolution of the DE density normalized to the present critical density ($\rho_{\rm crit,0}$) for different values of the coupling parameter $\xi$, including the case $\xi = 0$, which corresponds to the non-interacting scenario~\cite{Li:2019yem,Pan:2019hac}. This plot is particularly interesting, as it clearly demonstrates the late-time emergence of DE. In the right panel of Fig.~\ref{fig:energy-density}, we display the evolution of the density parameters of the dark components, namely $\Omega_{\rm DE} = \rho_{\rm DE}/\rho_{\rm crit}$ and $\Omega_{\rm CDM} = \rho_{\rm DM}/\rho_{\rm crit}$, for different values of $\xi$ ranging from negative to positive.

The evolution of the effective EoS parameters is shown in Fig.~\ref{fig:eff-eos}. From the left panel of Fig.~\ref{fig:eff-eos}, we observe that as $\xi$ decreases, the effective EoS of DE, $w_{\rm DE}^{\rm eff}$, approaches the boundary $-1$. Thus, for decreasing $\xi$, it is possible to realize a crossing from the phantom regime to the quintessential one, as recently suggested by DESI~\cite{DESI:2025zgx}. The evolution of $w_{\rm DM}^{\rm eff}$, shown in the right panel of Fig.~\ref{fig:eff-eos}, is simpler: for $\xi < 0$, one finds negative values of $w_{\rm DM}^{\rm eff}$.

\begin{figure*}
\includegraphics[width=0.47\textwidth]{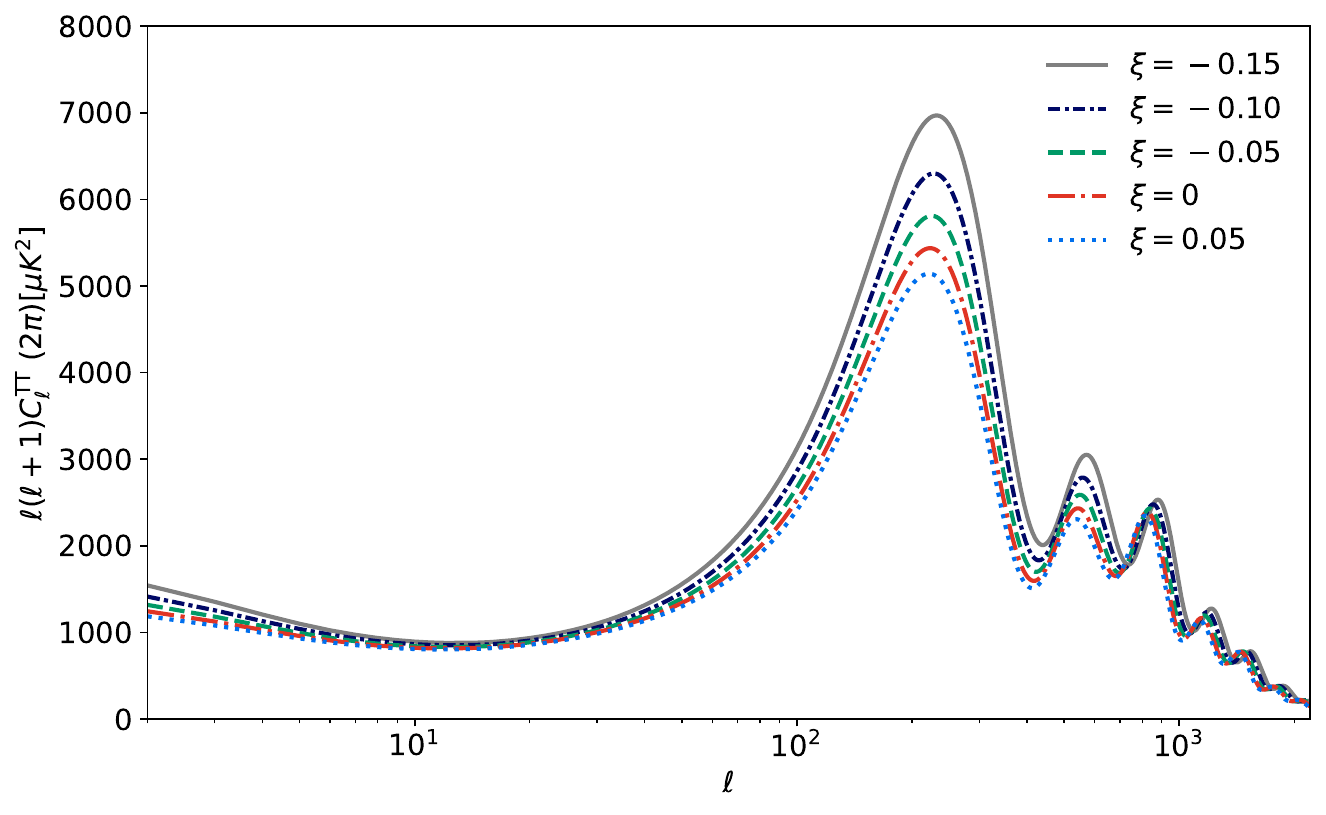}
\includegraphics[width=0.47\textwidth]{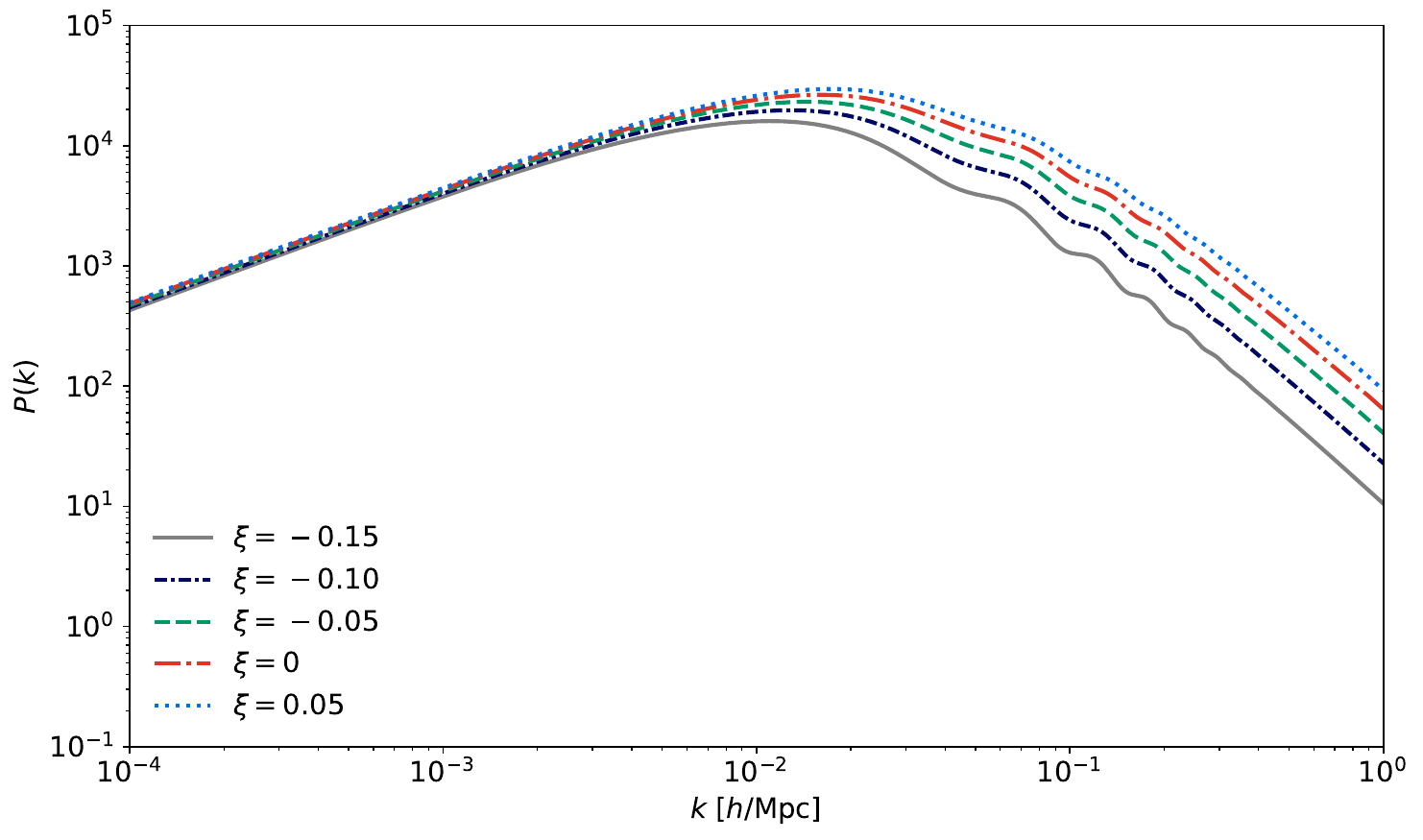}
\caption{Effects of different values of the coupling parameter in the \textbf{IPEDE} scenario on the CMB TT power spectrum (left) and the matter power spectrum (right). The remaining cosmological parameters ($\Omega_bh^2$, $\Omega_ch^2$ and $H_0$) are fixed to the mean values obtained from the CMB+DESI+PantheonPlus combination (see Table~\ref{tab:iPEDE}).  }
\label{fig:CMB-spectra}
\end{figure*}

In Fig.~\ref{fig:CMB-spectra}, we show the CMB TT and matter power spectra for different values of the coupling parameter $\xi$. It is evident that deviations from $\xi=0$ lead to noticeable changes in the CMB spectrum. For instance, when $\xi$ increases in the positive direction (corresponding to energy transfer from DM to DE), the height of the first acoustic peak in the CMB TT spectrum increases compared to the case $\xi = 0$. Conversely, for negative values of $\xi$, the height of the first peak decreases. Furthermore, modifications in the CMB TT spectrum are also observed at low multipoles in the presence of a non-vanishing coupling between DE and DM. The effects of the coupling parameter are also visible in the matter power spectrum. We find that for $\xi < 0$, a suppression in the matter power spectrum is observed. This can be understood as follows: $\xi < 0$ corresponds to a transfer of energy from DE to DM, so that the DM energy density decreases more slowly than $a^{-3}$, which is the standard behaviour in the absence of interaction. As a result, matter-radiation equality is delayed, leading to a suppression in the matter power spectrum. 

\begin{figure*}
    \centering
    \includegraphics[width=0.48\textwidth]{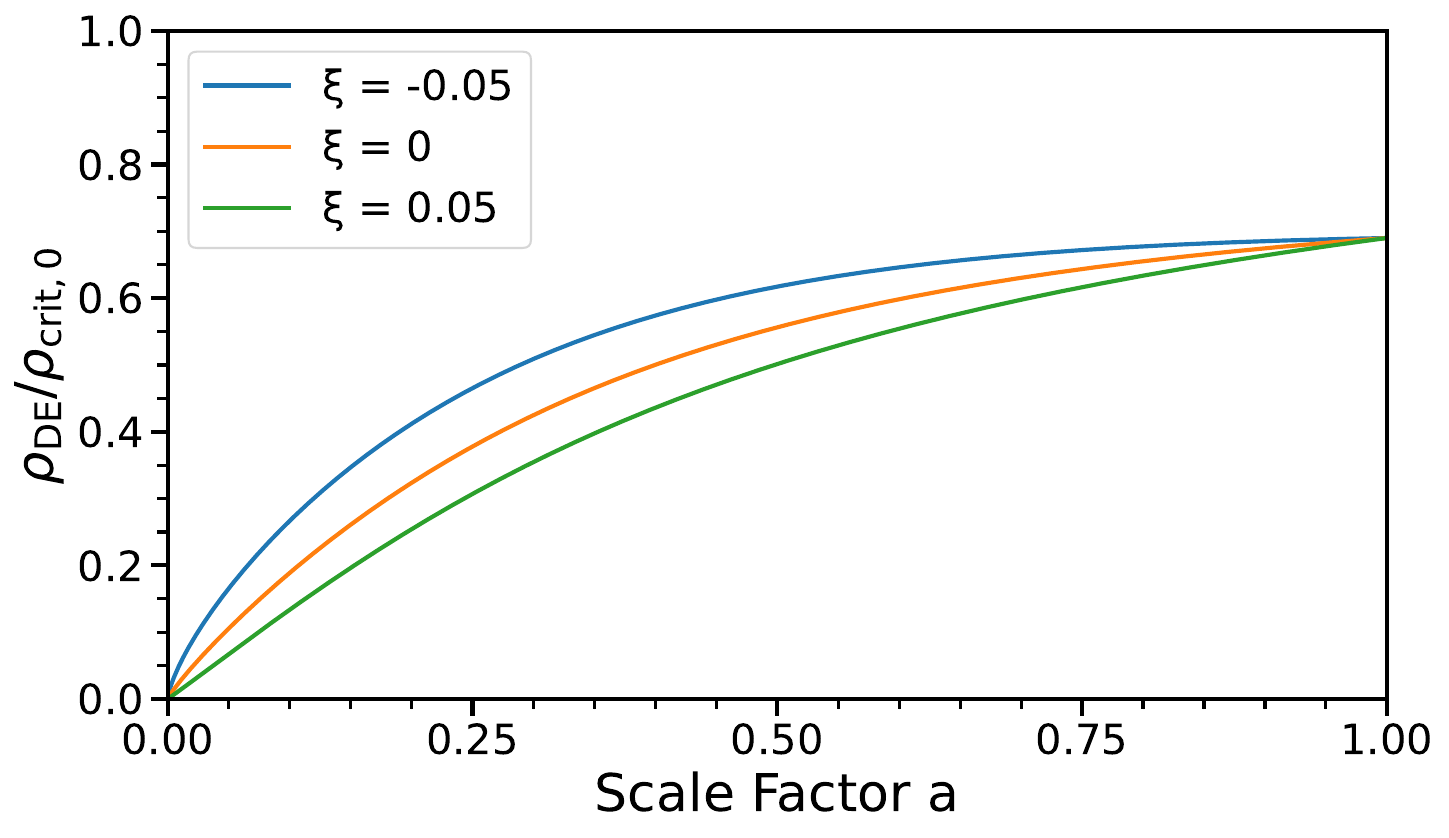}
    \includegraphics[width=0.48\textwidth]{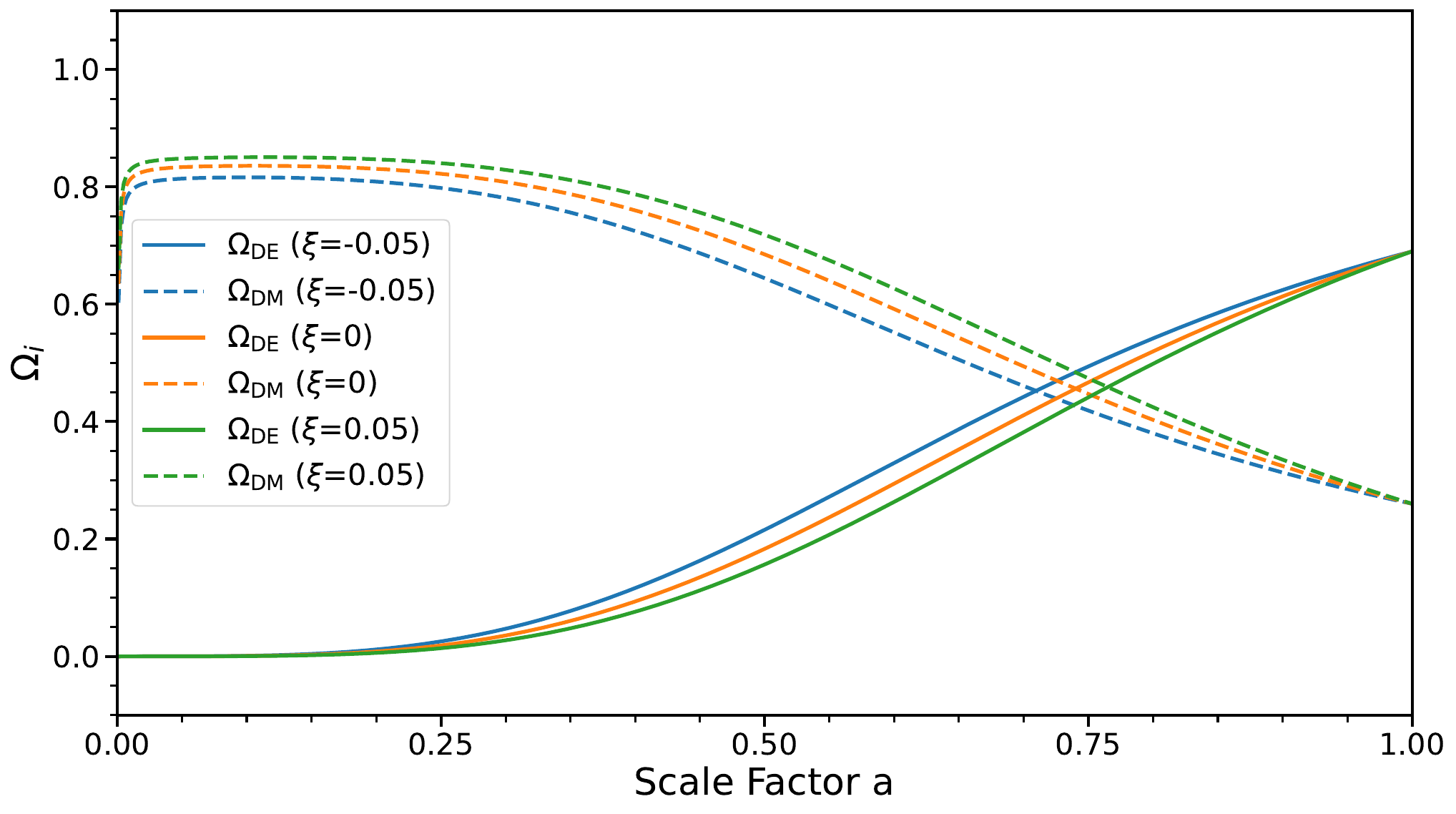}
    \caption{Evolution of $\rho_{\rm DE}/\rho_{\rm crit,0}$ (left panel) and of the DE and CDM density parameters, $\Omega_{\rm DE}$ and $\Omega_{\rm CDM}$ (right panel), for the {\bf IPEDE-gen} model, fixing $\alpha = 2$ and $a_t = 0.5$.}
    \label{fig:iPEDE-gen-density}
    \end{figure*}

\begin{figure*}
    \includegraphics[width=0.48\textwidth]{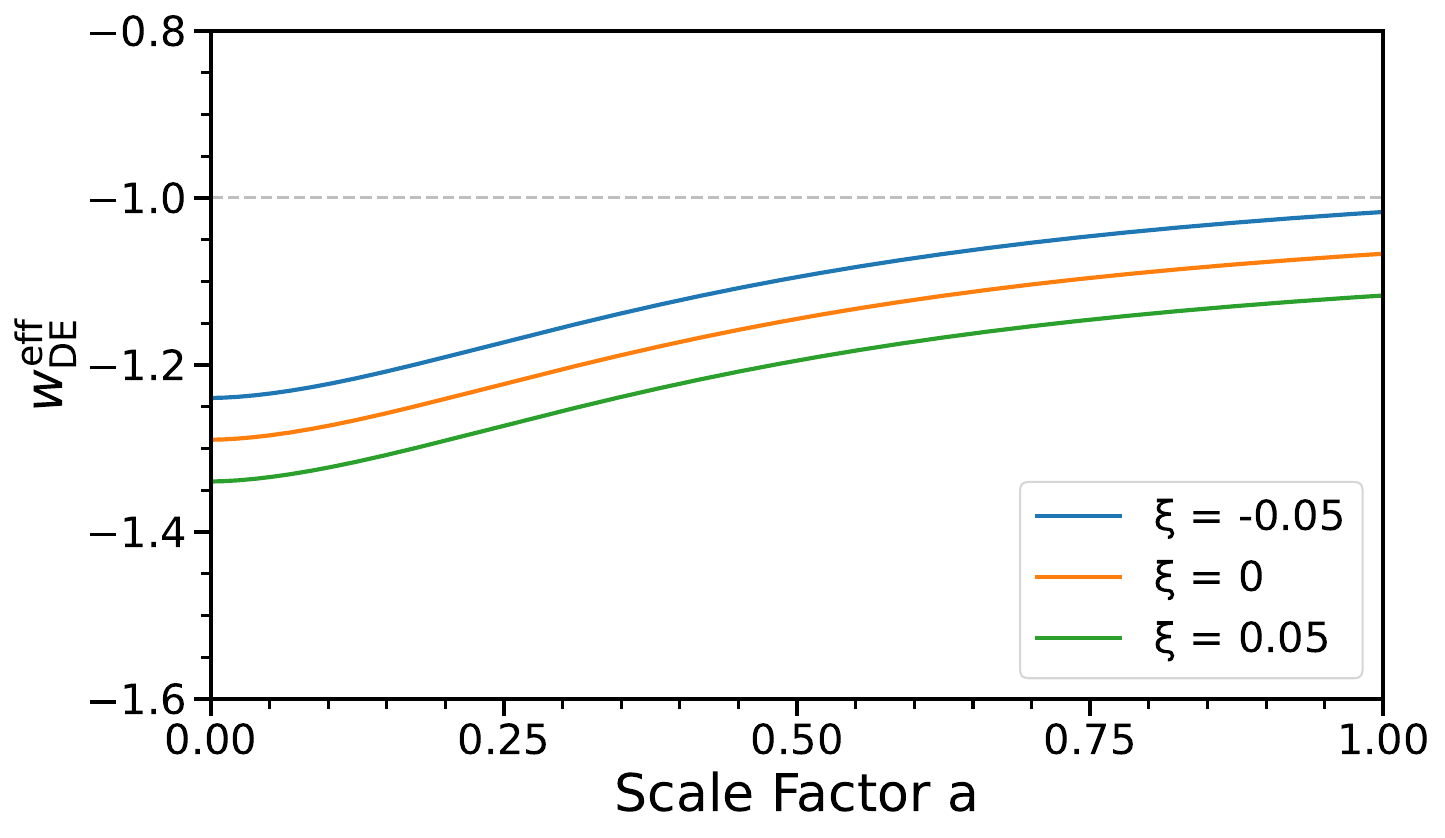}
    \includegraphics[width=0.48\textwidth]{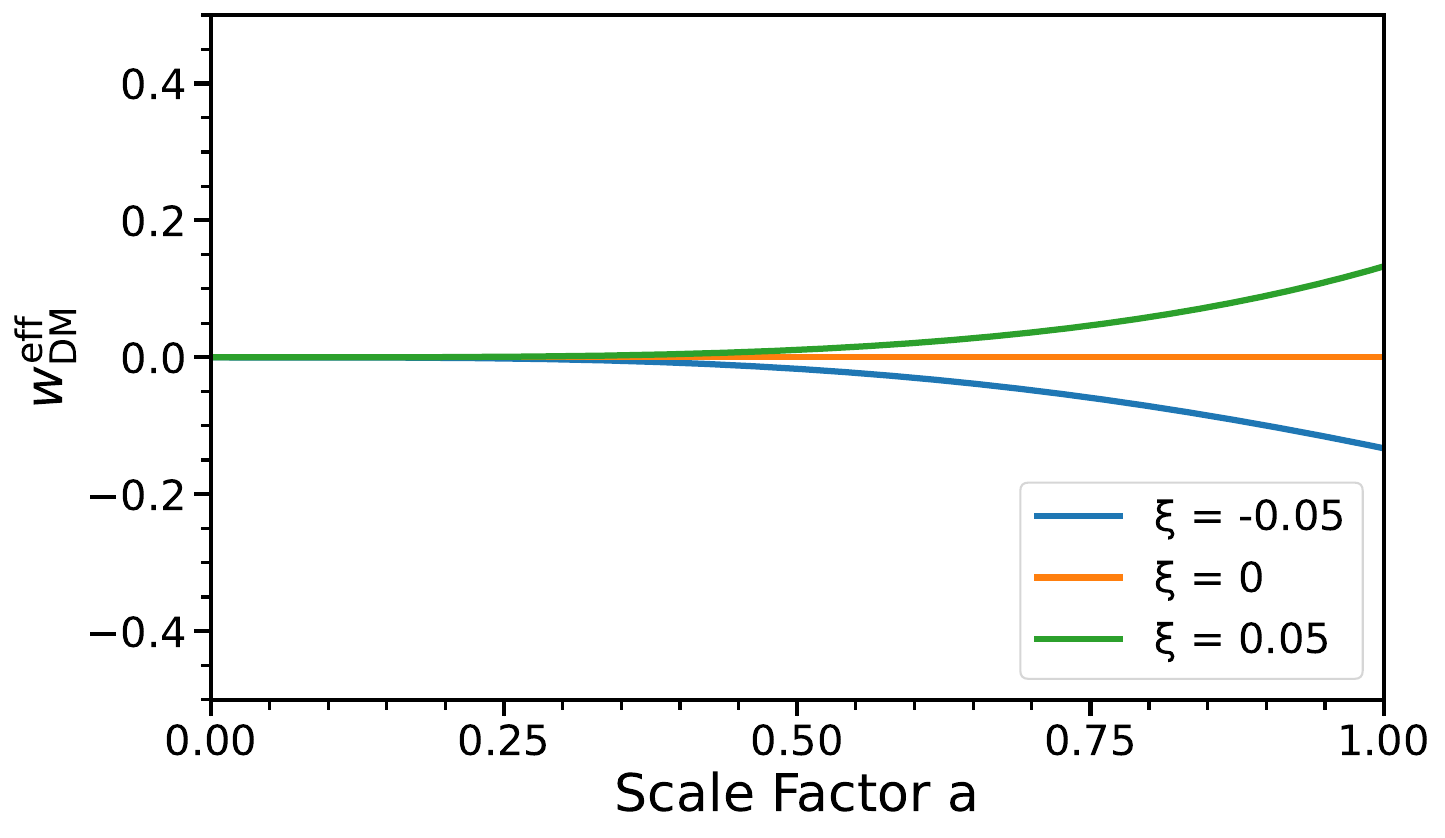}
    \caption{Evolution of the effective EoS parameters for different values of the coupling parameter $\xi$ in the {\bf IPEDE-gen} model, fixing $\alpha = 2$ and $a_t = 0.5$. The left panel shows $w_{\rm DE}^{\rm eff}$, while the right panel shows $w_{\rm DM}^{\rm eff}$.}
    \label{fig:iPEDE-gen-eff-eos}
\end{figure*}

We also consider a generalized EoS that includes both the epoch of transition (i.e.\ the epoch at which the emergent nature of DE turns on) and the speed of the transition. A possible generalization of Eq.~(\ref{eos}) is given by
\begin{align}\label{gen-eos}
w_{\rm DE}(a) = -1 - \frac{1}{3 \ln 10} \Bigl[ 1 - \tanh \bigl(\alpha \log_{10} (a/a_t) \bigr) \Bigr],
\end{align}
where $a_t$ is the scale factor corresponding to the epoch of transition and $\alpha >0$ determines the speed of the transition.  Although the generalized model extends the emergent DE model (\ref{eos}) through the introduction of two free parameters, however, within this generalized framework, we still notice that, $w_{\rm DE}(a) \sim -1.29$ as $a \rightarrow 0$, $w_{\rm DE}(a) = -1 - 1/(3 \ln 10)$ at the present epoch and $w_{\rm DE} (a)$ asymptotically approaches $-1$ in the far future. This nature remains similar to the earlier emergent DE model (\ref{eos}) because the modification does not affect qualitative nature of the DE EoS. A more general emergent DE model can treat the early-time value of the DE EoS as a free parameter and this may result in  a three-parameter emergent DE model. In the present work, however, we restrict our analysis to the above two-parameter extension of the original emergent DE model and investigate the DE-DM interaction.
We label the resulting interacting scenario driven by (\ref{gen-eos}) as {\bf IPEDE-gen}. 
We would like to  note here that even if $a_t \ll 1$ corresponding to a very early transition, this model does not necessarily recover $\Lambda$CDM as a limiting case, despite the fact that $w_{\rm DE} (a) \rightarrow -1$ in an asymptotic way. The reason is that the present framework incorporates an interaction between the dark fluids which therefore modifies the expansion history at the background and perturbative levels; and unless one sets $\xi =0$, $\Lambda$CDM is never 
recovered. 
As in the previous model, in the left panel of Fig.~\ref{fig:iPEDE-gen-density} we show the evolution of the DE density normalized to the present critical density for different values of the coupling parameter, fixing $\alpha = 2$ and $a_t = 0.5$. This plot clearly exhibits the emergent nature of DE. The evolution of the density parameters of the dark components (right panel of Fig.~\ref{fig:iPEDE-gen-density}), obtained for different values of $\xi$ (with $\alpha = 2$ and $a_t = 0.5$), is almost identical to the right panel of Fig.~\ref{fig:energy-density}. 
We now consider the evolution of the effective EoS parameters. In Fig.~\ref{fig:iPEDE-gen-eff-eos} we show the evolution of $w_{\rm DE}^{\rm eff}$ and $w_{\rm DM}^{\rm eff}$ for different values of the coupling parameter $\xi$, again fixing $\alpha = 2$ and $a_t = 0.5$. We observe that as $\xi$ decreases, $w_{\rm DE}^{\rm eff}$ approaches the boundary $-1$, similarly to the previous case, although here it lies even closer to this boundary. Therefore, it is evident that for decreasing $\xi$, a transition from the phantom regime to the quintessence regime can occur. The evolution of $w_{\rm DM}^{\rm eff}$, shown in the right panel of Fig.~\ref{fig:iPEDE-gen-eff-eos}, is also similar to the previous case, i.e.\ for $\xi < 0$, negative values of $w_{\rm DM}^{\rm eff}$ are obtained.

\begin{figure*}
    \centering
    \includegraphics[width=0.32\textwidth]{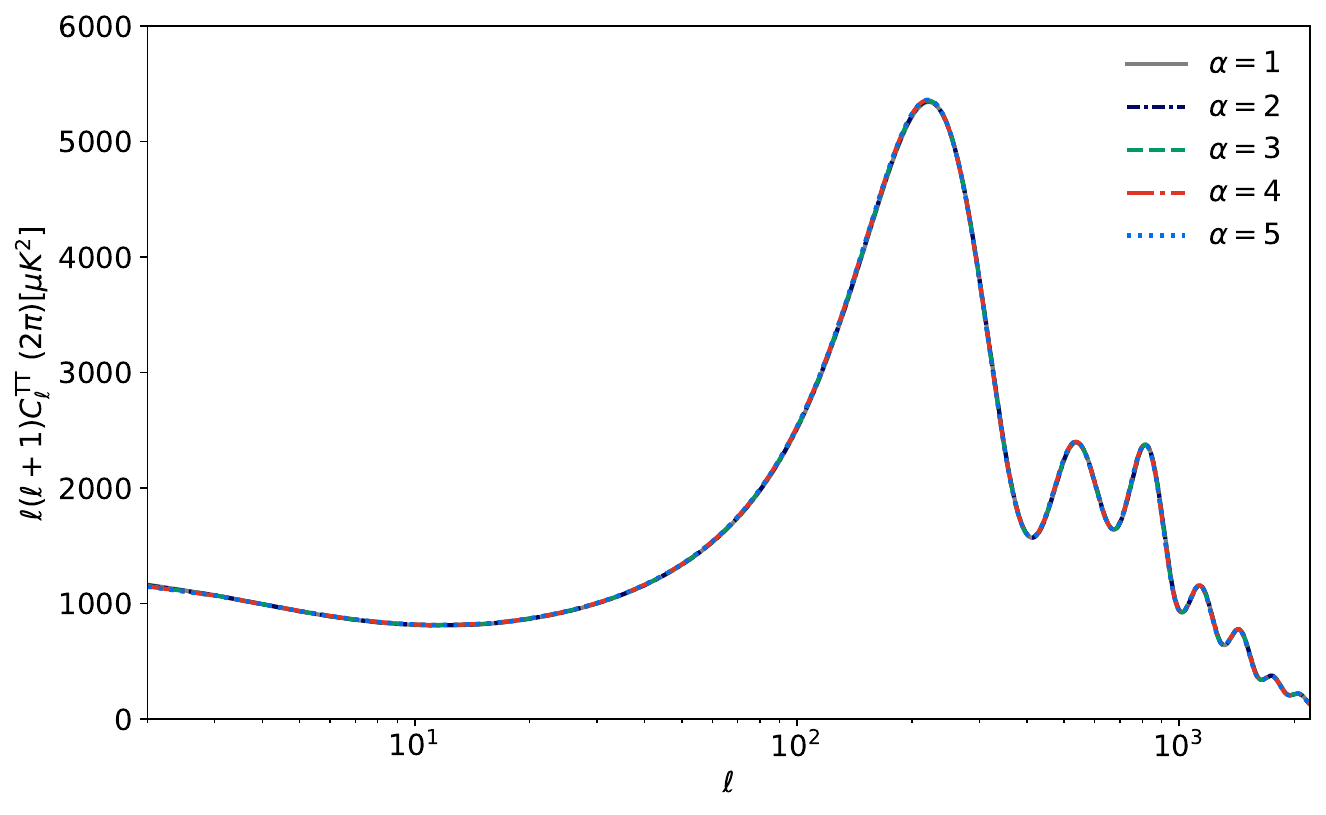}
    \includegraphics[width=0.32\textwidth]{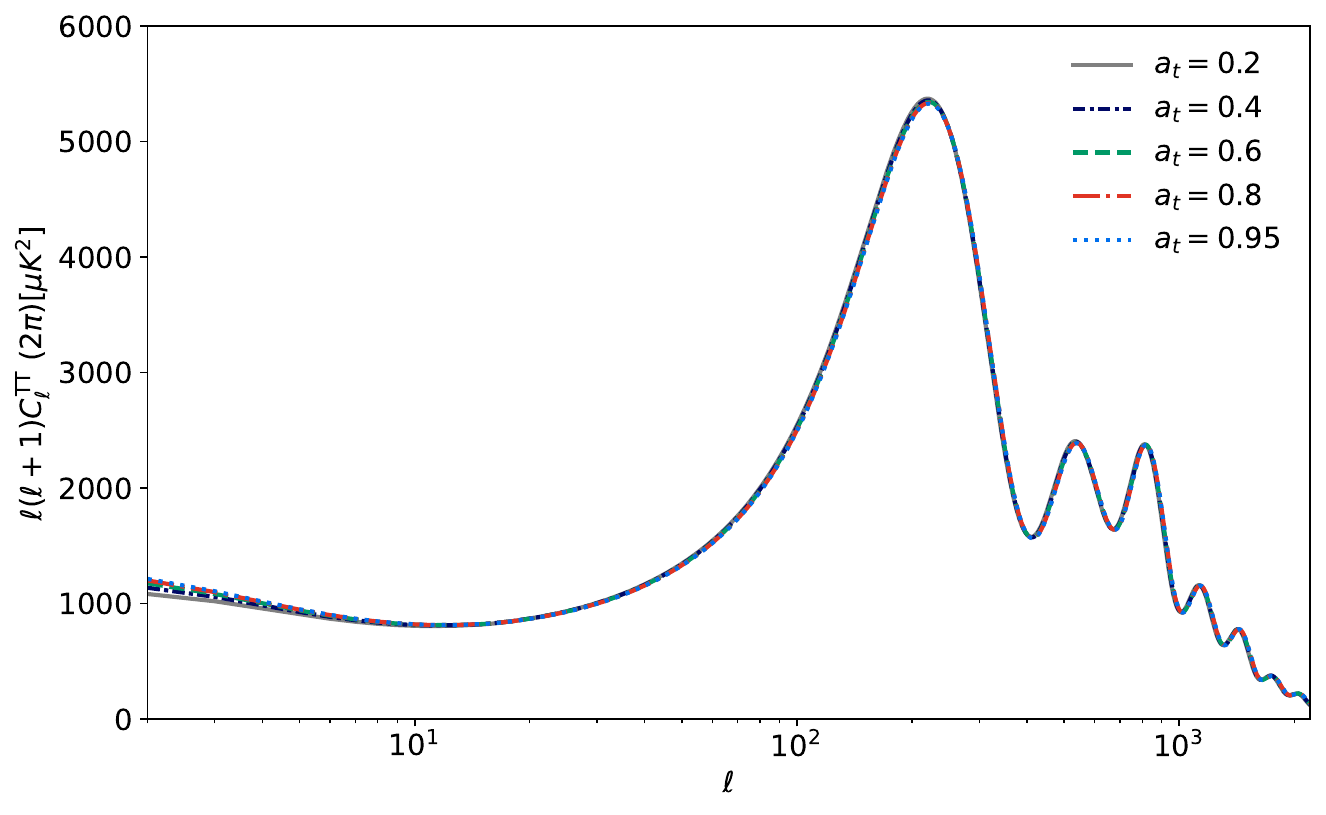}
    \includegraphics[width=0.32\textwidth]{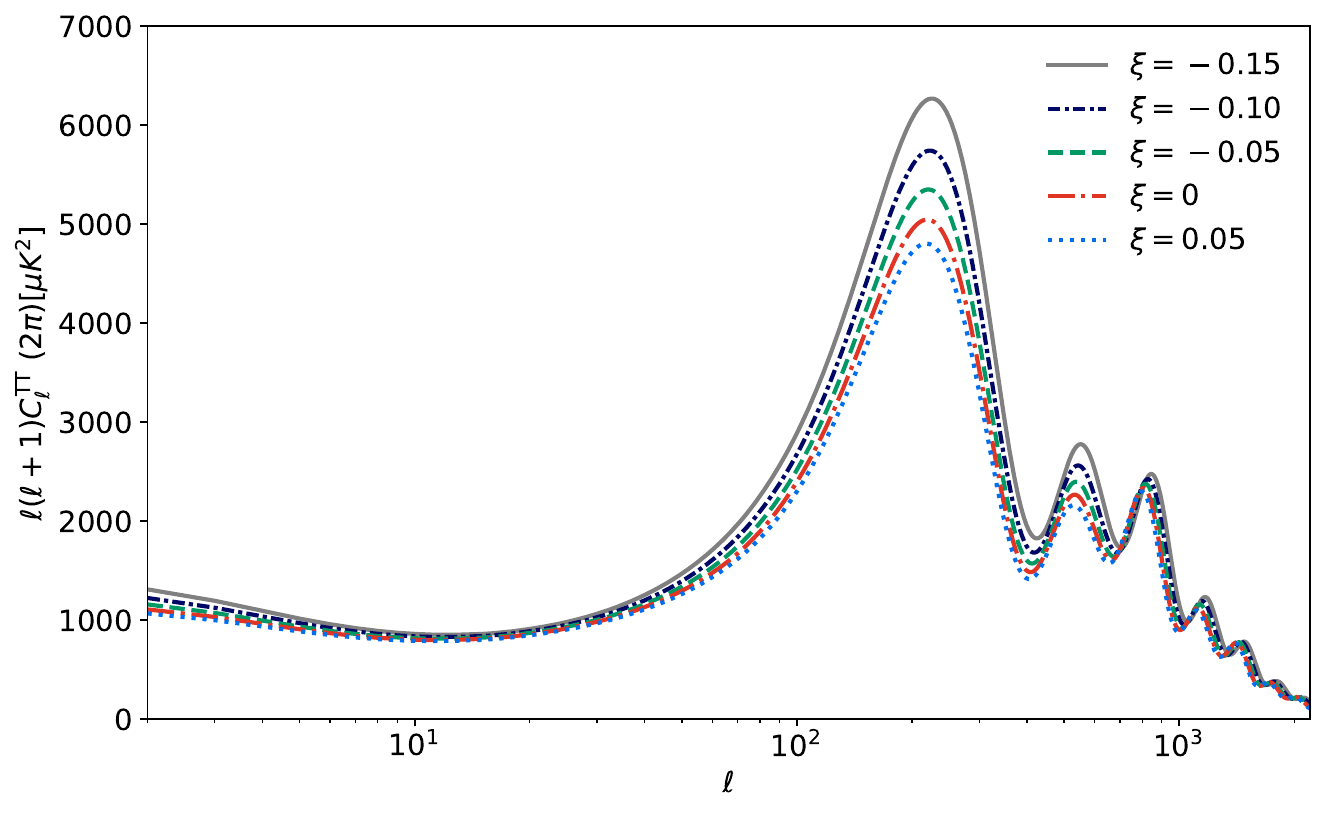}
    \includegraphics[width=0.32\textwidth]{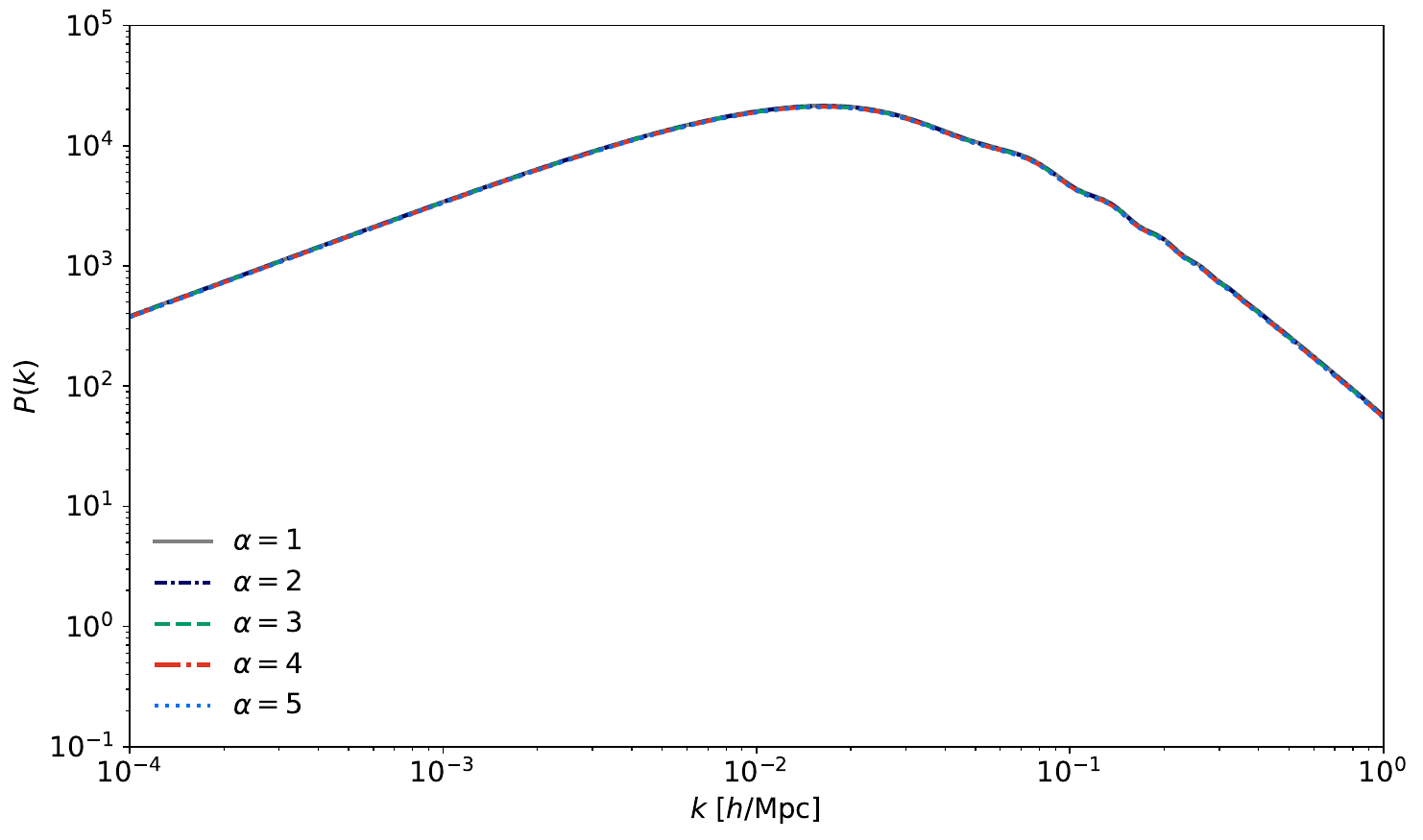}
    \includegraphics[width=0.32\textwidth]{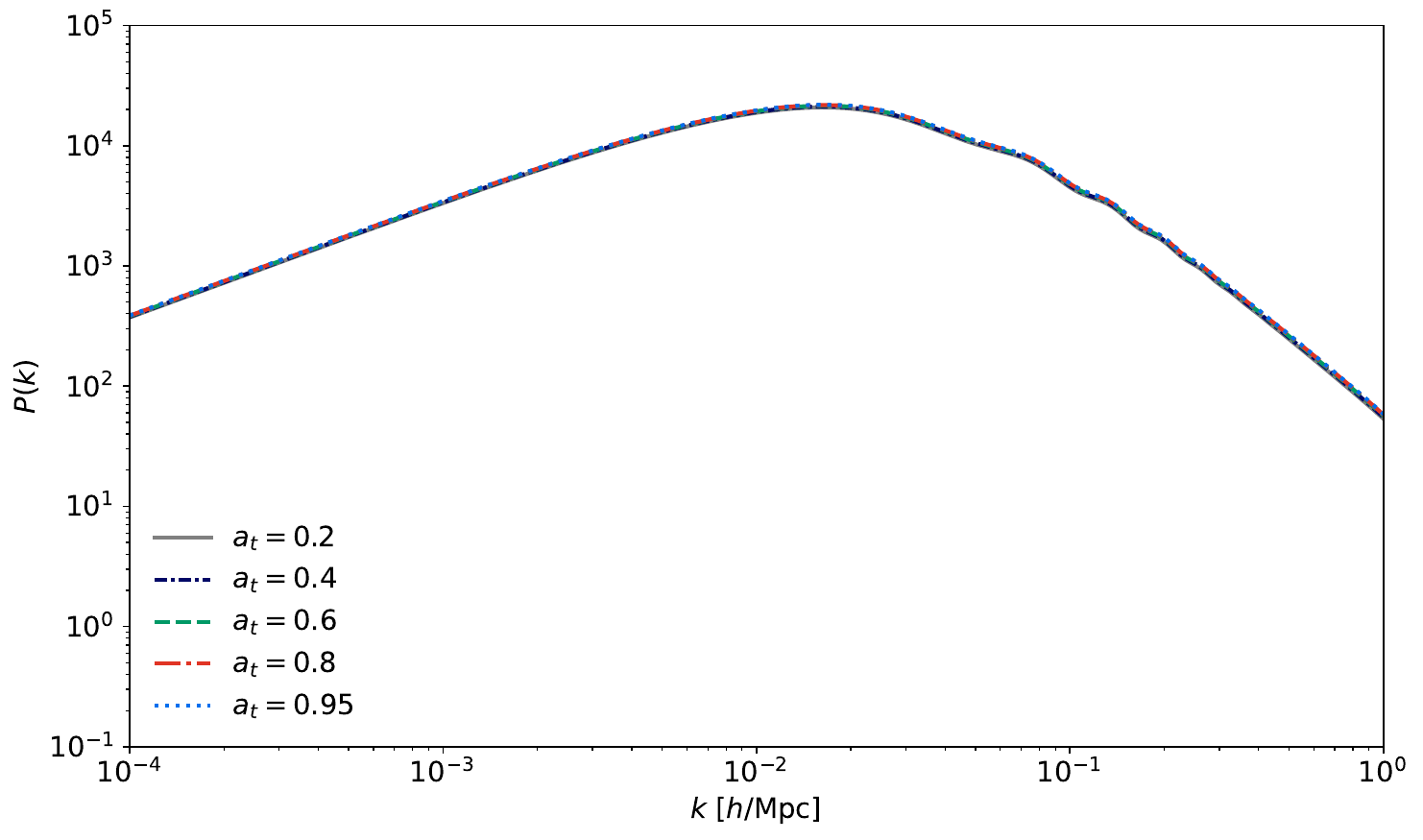}
    \includegraphics[width=0.32\textwidth]{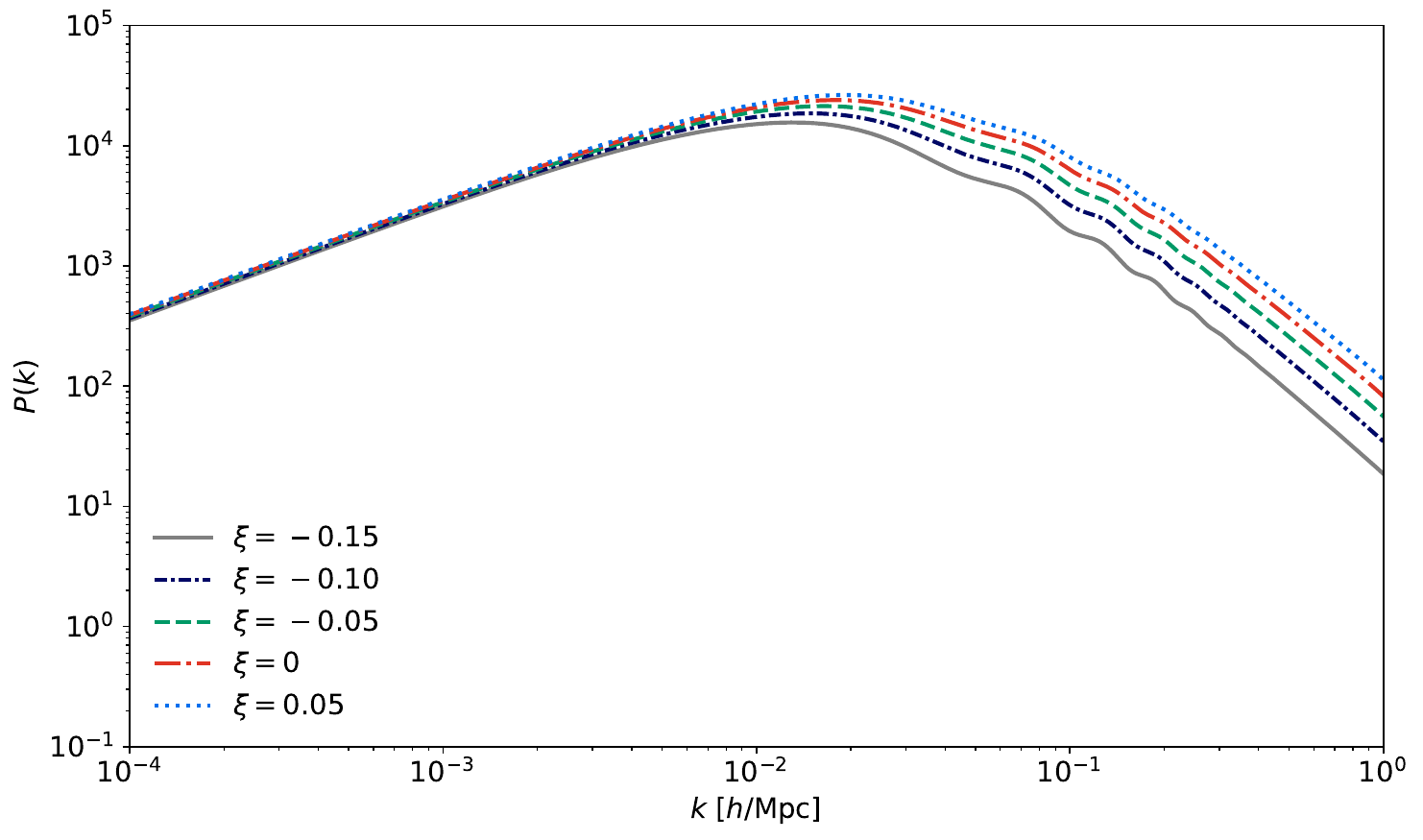}
    \caption{CMB TT power spectrum (upper panel) and matter power spectrum (lower panel) for the {\bf IPEDE-gen} scenario. When one parameter is varied, the other two are kept fixed: {\it (i)} when varying $\alpha$ (upper-left and lower-left panels), we fix $a_t = 0.5$ and $\xi = -0.05$; {\it (ii)} when varying $a_t$ (upper-middle and lower-middle panels), we fix $\alpha = 2$ and $\xi = -0.05$; and {\it (iii)} when varying $\xi$ (upper-right and lower-right panels), we fix $a_t = 0.5$ and $\alpha = 2$. The remaining cosmological parameters ($\Omega_{\rm b} h^2, \Omega_{\rm c} h^2, H_0$) are fixed to the mean values obtained from the CMB+DESI+PantheonPlus combination (see Table~\ref{tab:iPEDE-gen}).}
    \label{fig:iPEDE-gen-spectra}
\end{figure*}

In a similar fashion, in Fig.~\ref{fig:iPEDE-gen-spectra} we show the CMB TT and matter power spectra for: (i) variations in $\alpha$ (left panels of Fig.~\ref{fig:iPEDE-gen-spectra}), (ii) variations in $a_t$ (middle panels of Fig.~\ref{fig:iPEDE-gen-spectra}), and (iii) variations in $\xi$ (right panels of Fig.~\ref{fig:iPEDE-gen-spectra}). We find that variations in $\alpha$ do not affect the CMB spectra in any region, indicating that $\alpha$ is expected to remain largely unconstrained by CMB data. 
When varying $a_t$, mild changes are observed in the low-multipole region of the CMB spectra. The most significant effects on both the CMB and matter power spectra are found when varying $\xi$. A similar behaviour was observed in Fig.~\ref{fig:CMB-spectra}.


\begin{table}
	\begin{center}
		\renewcommand{\arraystretch}{1.4}
		\begin{tabular}{|c@{\hspace{1 cm}}|@{\hspace{1 cm}} c|}
			\hline
			\textbf{Parameter}           & \textbf{Prior}\\
			\hline\hline
			$\Omega_{b} h^2$             & $[0.005,0.1]$ \\
			$\Omega_{c} h^2$            & $[0.001,0.99]$ \\
			$\tau$                       & $[0.01,0.8]$ \\
			$n_s$                        & $[0.8, 1.2]$ \\
			$\log[10^{10}A_{s}]$         & $[1.61,3.91]$ \\
			$100\theta_{MC}$             & $[0.5,10]$ \\ 
			$\xi$                        & $[-1, 1]$ \\
            $\alpha$            & $[0, 10]$\\
            $a_t$               & $[0, 1]$ \\
			
			\hline
		\end{tabular}
	\end{center}
	\caption{Priors adopted for the cosmological parameters varied independently in the statistical analysis.}
	\label{tab:priors}
\end{table}

\section{Observational data and methodology}
\label{sec-data} 

In order to constrain the parameters of the interacting emergent dark energy scenario, we consider combinations of the following cosmological datasets:

\begin{enumerate}
\item {\bf Cosmic Microwave Background Radiation}: We use Cosmic Microwave Background (CMB) data from Planck 2018~\cite{Planck:2018vyg,Planck:2019nip}. Specifically, we consider the CMB temperature and polarization angular power spectra {\it plikTTTEEE+lowl+lowE}.   

\item {\bf Baryon Acoustic Oscillations}: We employ baryon acoustic oscillation (BAO) distance measurements from several astronomical surveys, with particular emphasis on the recent DESI BAO DR2 data~\cite{DESI:2025zgx}. These include measurements from the clustering of galaxies, quasars, and the Lyman-$\alpha$ forest over a wide redshift range. The BAO signal provides constraints on a set of key geometrical quantities, such as the spherically averaged distance scale $D_V(z)/r_d$, the comoving angular diameter distance $D_M(z)/r_d$, the angular diameter distance $D_A(z)/r_d$, and the Hubble parameter $H(z)r_d$, where $r_d$ is the comoving sound horizon at the baryon drag epoch.  

\item {\bf PantheonPlus}: Type Ia supernovae (SNe~Ia) serve as precise cosmological distance indicators, providing strong constraints on the expansion history of the universe and the dark energy equation of state. We employ the Pantheon+ supernova dataset~\cite{Scolnic:2021amr}, a comprehensive compilation of 1701 light curves from 1550 spectroscopically confirmed SNe Ia, systematically collected from eighteen independent astronomical surveys.

\item {\bf Union3}: We also adopt the Union3 dataset~\cite{Rubin:2023ovl}, which contains 2087 cosmologically useful SNe Ia from 24 independent surveys.

\item {\bf DES-Dovekie}: Finally, we utilize a new re-analysis of the 5-year supernova survey data from the {\bf Dark Energy Survey} ({\bf DES-Dovekie})~\cite{DES:2025sig}.\footnote{At this point, we would like to remark that the PantheonPlus compilation used here predates the Dovekie photometric recalibration. As reported in Refs.~\cite{Popovic:2025glk,DES:2025sig}, the Dovekie photometric recalibration may affect the PantheonPlus sample and therefore with the new PantheonPlus sample the constraints may alter. However, a fully recalibrated PantheonPlus likelihood is not yet publicly available. In the current article we use the publicly available PantheonPlus dataset~\cite{Scolnic:2021amr}, and in addition to that we have also used the recent DES-Dovekie~\cite{DES:2025sig} and Union3 ~\cite{Rubin:2023ovl} compilations. In the following section, we compare the cosmological constraints obtained from the three SN~Ia compilations aiming to examine how different SN~Ia dataset affects the constraints.}

\end{enumerate}
We now describe the methodology related to the statistical analyses. We modified \texttt{CAMB}~\cite{Lewis:2002ah,Lewis:1999bs}, a freely available advanced cosmological package, and performed Markov Chain Monte Carlo (MCMC) analyses using \texttt{Cobaya}~\cite{2019ascl.soft10019T,Torrado:2020dgo}, a publicly available sampler. The MCMC chains were run until convergence according to the Gelman--Rubin criterion~\cite{Gelman:1992zz}, requiring $R-1 < 0.01$. The model contains seven free parameters: one coupling parameter, $\xi$, and six standard $\Lambda$CDM parameters, namely, $\Omega_{\rm b} h^2$ (physical baryon density), $\Omega_{\rm c} h^2$ (physical CDM density), $\tau$ (optical depth to reionization), $\theta_{MC}$ (angular scale of the sound horizon at recombination), $A_{\mathrm{s}}$ (amplitude of the primordial scalar perturbations), and $n_{\mathrm{s}}$ (scalar spectral index).  
In Table~\ref{tab:priors}, we show the flat priors assumed for the seven free parameters of this interacting model.

For model selection, we also evaluate the Bayesian evidence. For a model $\mathcal{M}_i$ with parameter vector $\Theta$, the evidence is $B_i = \int \mathcal{L}(D|\Theta,\mathcal{M}_i)\,\pi(\Theta|\mathcal{M}_i)\,d\Theta$, where $\mathcal{L}$ is the likelihood of the data $D$ and $\pi$ the prior. Model comparison proceeds via the Bayes factor $B_{ij}=B_i/B_j$ ($B_i$ is the interacting scenario and $B_j$ is the reference model), commonly reported as the relative log-evidence $\ln B_{ij} = \ln B_i - \ln B_j$, with positive values favoring model $i$ over $j$. We interpret $\ln B_{ij}$ using the revised Jeffreys’ scale~\cite{Kass:1995loi}: $[0,1]$ \textit{inconclusive}, $[1,2.5]$ \textit{weak}, $[2.5,5]$ \textit{moderate}, $[5,10]$ \textit{strong}, and $>10$ \textit{very strong} evidence. In this article, we consider two reference models: $\Lambda$CDM and the Chevallier-Polarski-Linder (CPL) model~\cite{Chevallier:2000qy,Linder:2002et}. 
In practice, we compute $\ln B = \ln B_{\text{Model}} - \ln B_{\Lambda\text{CDM}~{\rm or~ CPL}}$ with \texttt{MCEvidence}~\cite{Heavens:2017hkr,Heavens:2017afc}. Alongside this, we use $\Delta \chi^2_{\rm min} = \chi^2_{\rm min}(\text{Model}) - \chi^2_{\rm min}(\Lambda\text{CDM}~{\rm or~ CPL})$. 
These statistics will be used in Section~\ref{sec-results} to assess the performance of the {\bf IPEDE} and  {\bf IPEDE-gen} models relative to $\Lambda$CDM and CPL. Under the above definitions,  $\Delta\chi^2_{\rm min} < 0$ indicates that the model is preferred over the reference model as per the chi-squared statistic, and $\ln B_{ij} >0$ indicates that the model is favored as per Bayesian evidence analysis. 
\begingroup
\begin{center}
	\begin{table*}
		\scalebox{0.85}{
			\begin{tabular}{ccccccccc}
				\hline\hline
				Parameters & CMB & CMB+DESI & CMB+DESI+PantheonPlus & CMB+DESI+Union3 & CMB+DESI+DES-Dovekie\\ \hline
				
				$\Omega_b h^2$ & 
				$0.02231_{-0.00016-0.00029}^{+0.00016+0.00031}$ & 
				$0.02242_{-0.00013-0.00025}^{+0.00013+0.00026}$ & 
				$0.02246_{-0.00013-0.00026}^{+0.00013+0.00026}$ & 
				$0.02246_{-0.00013-0.00026}^{+0.00013+0.00026}$ & 
				$0.02246_{-0.00013-0.00024}^{+0.00012+0.00025}$ \\
				
				$\Omega_c h^2$ & 
				$0.1197_{-0.0015-0.0028}^{+0.0015+0.0029}$ & 
				$0.11799_{-0.00083-0.0017}^{+0.00084+0.0017}$ & 
				$0.11729_{-0.00080-0.0015}^{+0.00078+0.0016}$ & 
				$0.11732_{-0.00081-0.0016}^{+0.00081+0.0016}$ & 
				$0.11729_{-0.00074-0.0015}^{+0.00076+0.0015}$ \\
				
				$100\theta_{MC}$ & 
				$1.04065_{-0.00033-0.00065}^{+0.00033+0.00063}$ & 
				$1.04089_{-0.00028-0.00055}^{+0.00028+0.00055}$ & 
				$1.04097_{-0.00028-0.00055}^{+0.00028+0.00055}$ & 
				$1.04098_{-0.00028-0.00055}^{+0.00028+0.00054}$ & 
				$1.04100_{-0.00027-0.00055}^{+0.00027+0.00053}$ \\
				
				$\tau$ & 
				$0.0536_{-0.0079-0.016}^{+0.0079+0.016}$ & 
				$0.0581_{-0.0079-0.015}^{+0.0079+0.016}$ & 
				$0.0603_{-0.0072-0.015}^{+0.0089+0.017}$ & 
				$0.0604_{-0.0077-0.016}^{+0.0087+0.017}$ & 
				$0.0598_{-0.0088-0.015}^{+0.0073+0.018}$ \\
				
				$n_s$ & 
				$0.9729_{-0.0044-0.0088}^{+0.0044+0.0084}$ & 
				$0.9765_{-0.0035-0.0071}^{+0.0035+0.0068}$ & 
				$0.9779_{-0.0035-0.0068}^{+0.0035+0.0069}$ & 
				$0.9778_{-0.0034-0.0067}^{+0.0034+0.0066}$ & 
				$0.9779_{-0.0034-0.0066}^{+0.0035+0.0065}$ \\
				
				${\rm{ln}}(10^{10} A_s)$ & 
				$3.053_{-0.016-0.032}^{+0.016+0.032}$ & 
				$3.059_{-0.016-0.032}^{+0.016+0.033}$ & 
				$3.061_{-0.015-0.032}^{+0.018+0.036}$ & 
				$3.061_{-0.017-0.033}^{+0.017+0.034}$ & 
				$3.060_{-0.018-0.033}^{+0.015+0.037}$ \\
				
				$\xi$ & 
				$0.023_{-0.070-0.19}^{+0.13+0.17}$ & 
				$-0.106_{-0.039-0.074}^{+0.039+0.079}$ & 
				$-0.164_{-0.024-0.045}^{+0.024+0.048}$ & 
				$-0.163_{-0.028-0.053}^{+0.028+0.055}$ & 
				$-0.172_{-0.021-0.041}^{+0.021+0.042}$ \\
				
				$\Omega_m$ & 
				$0.272_{-0.033-0.046}^{+0.018+0.056}$ & 
				$0.2921_{-0.0075-0.015}^{+0.0075+0.015}$ & 
				$0.3029_{-0.0052-0.010}^{+0.0052+0.010}$ & 
				$0.3028_{-0.0059-0.012}^{+0.0059+0.012}$ & 
				$0.3048_{-0.0048-0.009}^{+0.0047+0.010}$ \\
				
				$\sigma_8$ & 
				$0.91_{-0.21-0.30}^{+0.16+0.33}$ & 
				$0.695_{-0.041-0.093}^{+0.052+0.097}$ & 
				$0.629_{-0.024-0.051}^{+0.028+0.052}$ & 
				$0.630_{-0.027-0.058}^{+0.032+0.059}$ & 
				$0.621_{-0.023-0.044}^{+0.023+0.046}$ \\
				
				$H_0$ [Km/s/Mpc] & 
				$72.6_{-2.8-6.6}^{+4.0+6.0}$ & 
				$69.52_{-0.95-1.9}^{+0.95+1.9}$ & 
				$68.09_{-0.59-1.1}^{+0.59+1.2}$ & 
				$68.11_{-0.69-1.3}^{+0.69+1.4}$ & 
				$67.88_{-0.53-1.05}^{+0.52+1.01}$ \\
				
				$S_8$ & 
				$0.86_{-0.15-0.21}^{+0.11+0.24}$ & 
				$0.685_{-0.035-0.078}^{+0.044+0.082}$ & 
				$0.632_{-0.022-0.046}^{+0.025+0.047}$ & 
				$0.633_{-0.024-0.052}^{+0.028+0.052}$ & 
				$0.626_{-0.022-0.041}^{+0.021+0.043}$ \\
				
				$r_{\rm{drag}}$ [Mpc] & 
				$147.13_{-0.31-0.60}^{+0.31+0.61}$ & 
				$147.46_{-0.21-0.43}^{+0.24+0.47}$ & 
				$147.61_{-0.22-0.43}^{+0.22+0.42}$ & 
				$147.60_{-0.22-0.43}^{+0.22+0.43}$ & 
				$147.61_{-0.21-0.41}^{+0.20+0.41}$ \\
                  \hline 
                $\Delta \chi^2_{\rm min}~({\rm w.r.t}~\text{$\Lambda$CDM})$ & 
                $1.4$ &
                $4.8$ &
                $6.3$ &
                $6.1$ &
                $5.5$\\
				$\ln B_{ij}~({\rm w.r.t}~\text{$\Lambda$CDM})$ & 
				$-3.4$ & 
				$-6.2$ & 
				$-7.2$ & 
				$-6.8$ & 
				$-5.6$ \\
                  \hline 
				$\Delta \chi^2_{\rm min}~{\rm (w.r.t ~CPL)}$ & 
				$0.1$ &
				$10.1$ &
				$11.8$ &
				$17.4$ &
				$14.1$\\
				$\ln B_{ij}~{\rm (w.r.t~CPL)}$ & 
				$-3.7$ & 
				$-7.1$ & 
				$-6.7$ & 
				$-10.1$ & 
				$-6.7$ \\			
				\hline                       
		\end{tabular}}                                                                
		\caption{68\% and 95\% CL constraints on the \textbf{IPEDE} scenario from CMB, CMB+DESI, CMB+DESI+PantheonPlus, CMB+DESI+Union3, and CMB+DESI+DES-Dovekie.
         }
		\label{tab:iPEDE}                   
	\end{table*}                                     
\end{center}
\endgroup
\begin{figure*}
	\includegraphics[width=0.9\textwidth]{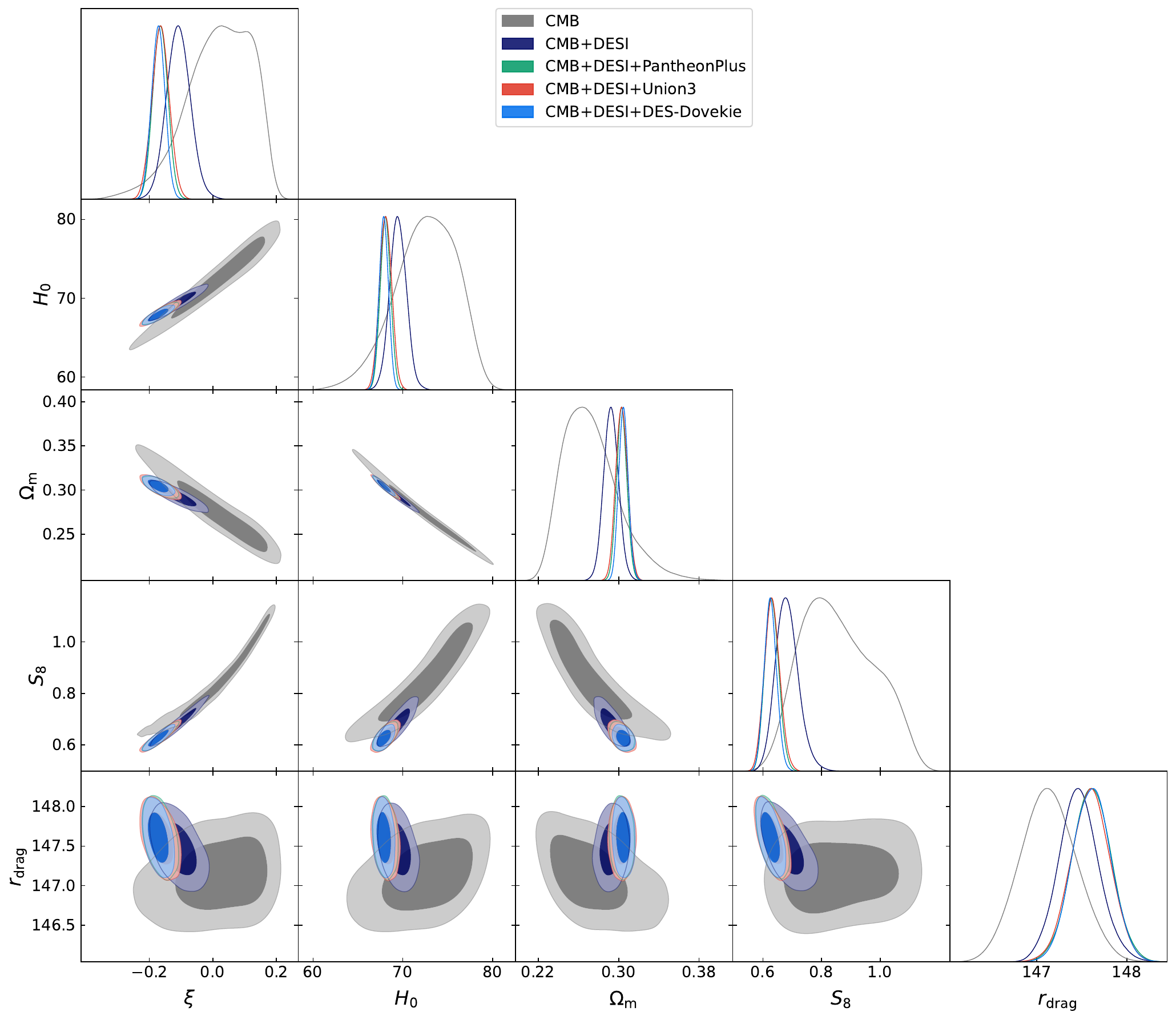}
	\caption{One-dimensional posterior distributions and two-dimensional joint contours for the most relevant parameters of the \textbf{IPEDE} scenario from different combinations of cosmological datasets. }
	\label{fig:IPEDE-1}
\end{figure*}

\begin{figure*}
	\includegraphics[width=0.45\textwidth]{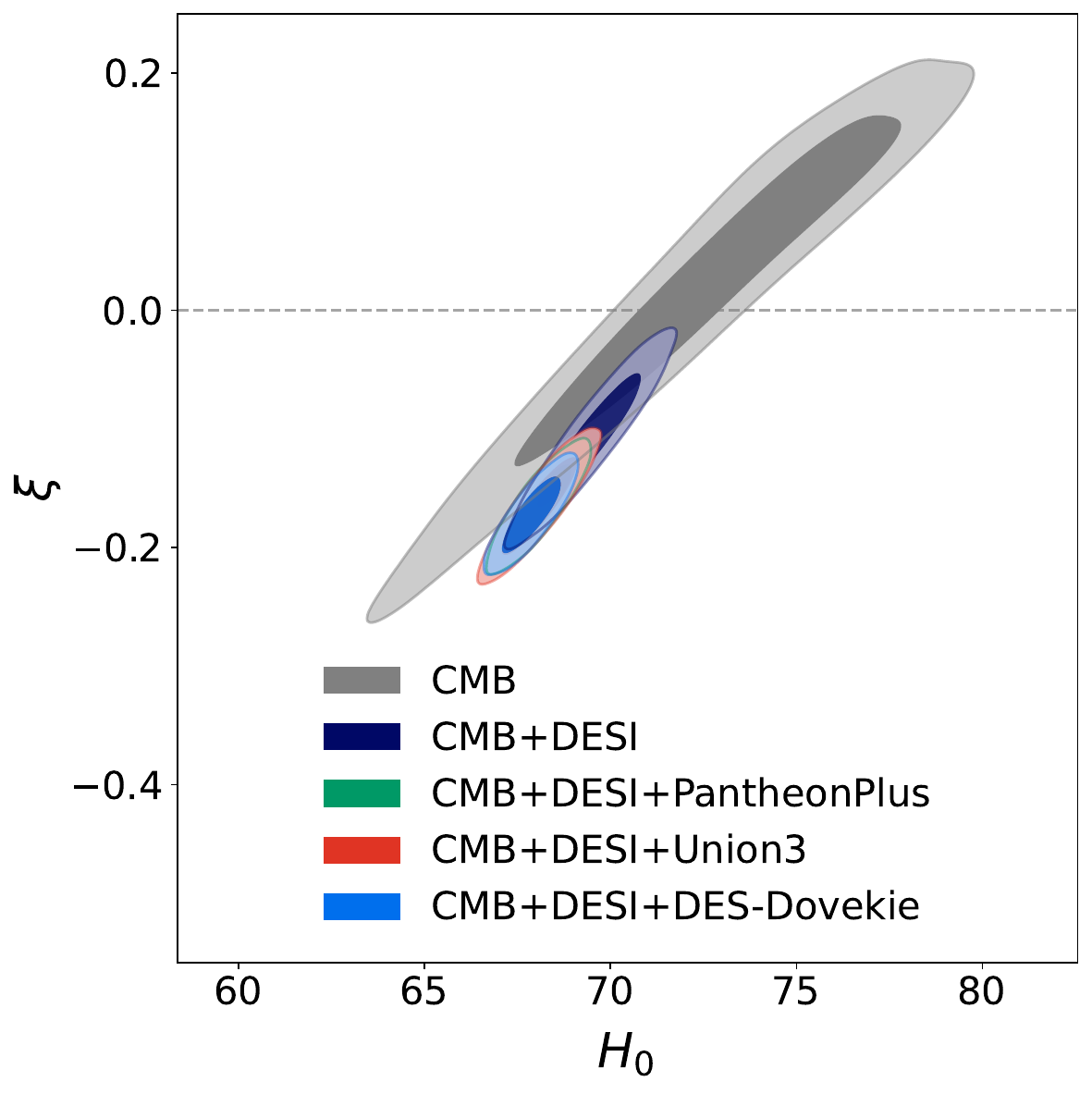}
	\includegraphics[width=0.45\textwidth]{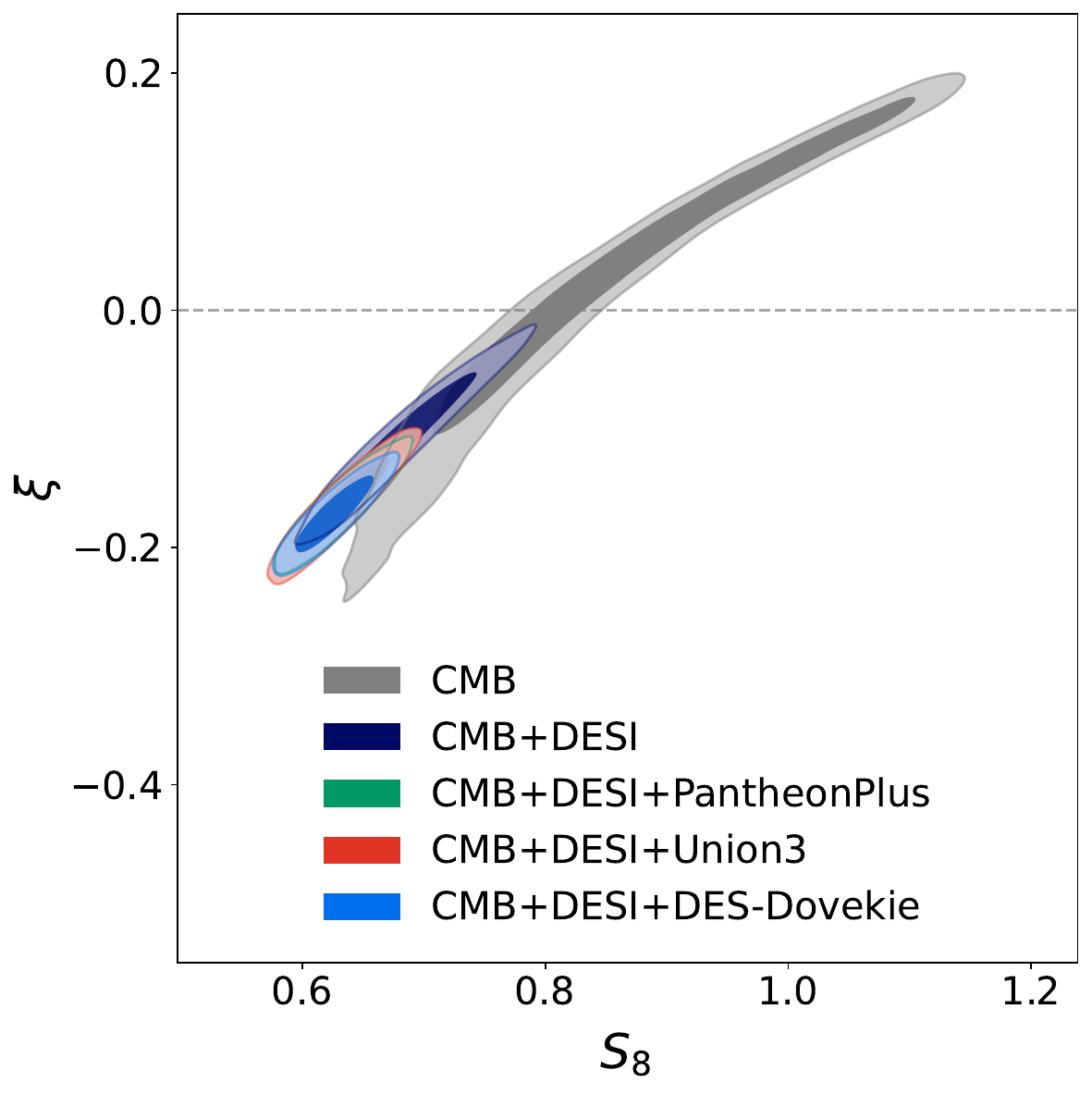}
	\caption{Left: constraints in the $(H_0, \xi)$ plane for various dataset combinations. Right: constraints in the $(\xi, S_8)$ plane for all datasets.  }
	\label{fig:H0-S8-xi}
\end{figure*}

\begin{figure*}
    \includegraphics[width=0.495\textwidth]{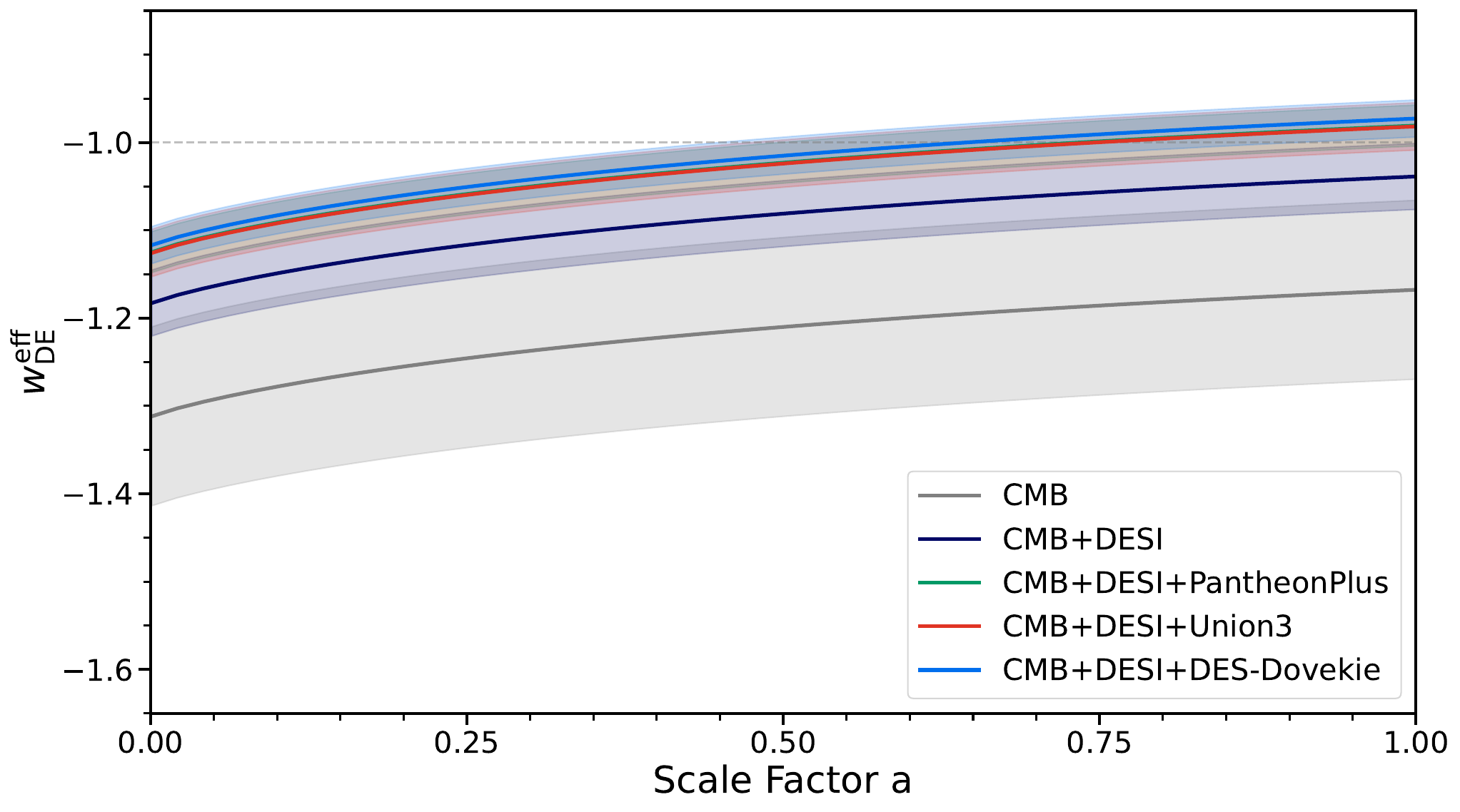}
    \includegraphics[width=0.495\textwidth]{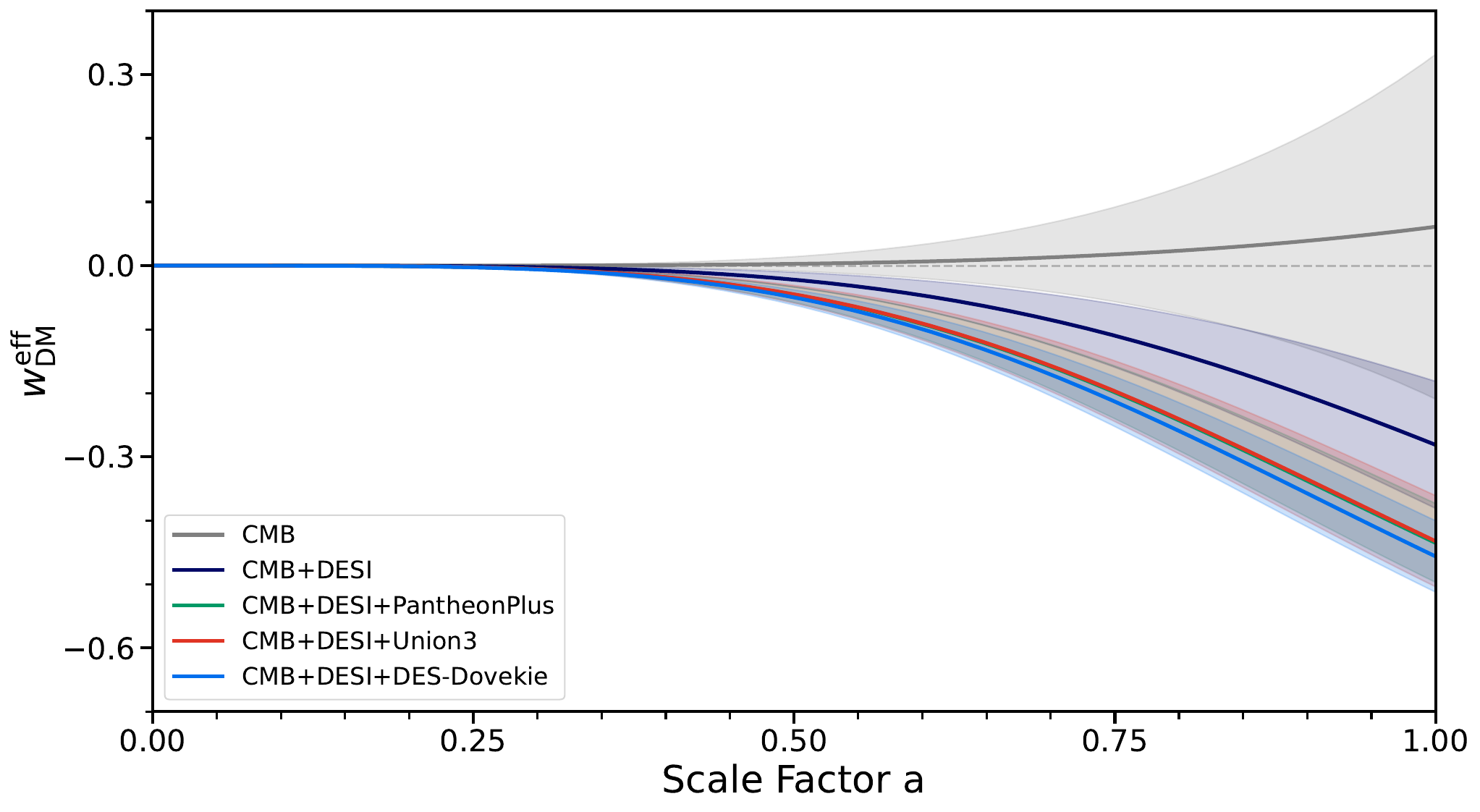}
    \caption{Evolution of the effective EoS parameters, including their 68\% CL credible regions, as a function of the scale factor for {\bf IPEDE} scenario. The left panel shows $w_{\rm DE}^{\rm eff}$, while the right panel shows $w_{\rm DM}^{\rm eff}$. In both plots, the curve obtained from CMB+DESI+PantheonPlus (green) is indistinguishable from that of CMB+DESI+Union3 (red), since the constraints from these combined datasets are nearly identical.  }
    \label{fig:eff-eos-data}
\end{figure*}

\section{Results and analyses}
\label{sec-results}

In this section we summarize the observational constraints on the \textbf{IPEDE} and {\bf IPEDE-gen} scenarios from the datasets described in Section~\ref{sec-data}. To assess statistical viability, we use the indicators introduced in Section~\ref{sec-data}, namely $\Delta \chi^2_{\rm min}$ and the logarithmic Bayes factor $\Delta \ln B_{ij}$ (with $\Delta \chi^2_{\rm min} > 0$ and $\Delta \ln B_{ij} < 0$ indicating preference for the reference $\Lambda$CDM model). In the following, we present the key results of this interacting scenario.

\subsection{IPEDE}

In  Table~\ref{tab:iPEDE} and in Figs.~\ref{fig:IPEDE-1} and~\ref{fig:H0-S8-xi} we summarize the main results of the interacting scenario. We start with the constraints from CMB alone (second column of Table~\ref{tab:iPEDE}). In this case, the data do not indicate any significant preference for a coupling between the dark sectors. Nevertheless, as is already well known for the non-interacting PEDE framework, CMB data alone lead to a higher value of the Hubble constant, yielding $H_0 = 72.6^{+4.0}_{-2.8}$ km/s/Mpc at 68\% CL. This alleviates the long-standing Hubble tension between Planck~\cite{Planck:2018vyg} and SH0ES~\cite{Riess:2021jrx} (see also the review~\cite{DiValentino:2021izs}). The parameter $S_8$ also takes a higher value in this framework. Overall, with CMB data alone the IPEDE scenario provides a reasonable solution to the $H_0$ tension, but not to the $S_8$ tension.

When DESI is combined with CMB, notable changes are found in the constraints of the interacting scenario. In particular, we now obtain evidence for a coupling between the dark fluids at more than 95\% CL: $\xi = -0.106^{+0.079}_{-0.074}$ (CMB+DESI). The Hubble constant remains relatively high, with $H_0 = 69.52^{+0.95}_{-0.95}$ km/s/Mpc at 68\% CL (CMB+DESI), compared to the Planck baseline assuming $\Lambda$CDM~\cite{Planck:2018vyg}. However, relative to the CMB-alone case in this interacting framework, the value of $H_0$ is slightly reduced. This reduction can be understood from the direction of the energy flow: here, $\xi < 0$ at more than 95\% CL, which implies a transfer of energy from DE to CDM. As a consequence, $\Omega_m$ increases compared to the CMB-alone case ($\Omega_m = 0.272^{+0.018}_{-0.033}$ at 68\% CL, CMB alone), and the value of $H_0$ is mildly lowered due to the well-known geometrical degeneracy between $\Omega_m$ and $H_0$, as illustrated in Fig.~\ref{fig:IPEDE-1}.  
Importantly, even with this reduction, the Hubble constant obtained with CMB+DESI still corresponds to a value that alleviates the $H_0$ tension to below the $3\sigma$ level when compared to local determinations from SH0ES~\cite{Riess:2021jrx}, while remaining consistent with the fact that PEDE-type models are known to push $H_0$ in the right direction.  
Alongside these changes, the most notable shift occurs for the clustering parameter $S_8$. With CMB+DESI we find $S_8 = 0.685^{+0.044}_{-0.035}$ at 68\% CL, which is significantly lower than the value for the CMB-alone case ($S_8 = 0.86^{+0.11}_{-0.15}$ at 68\% CL, CMB alone), and also lower than the value inferred in the Planck $\Lambda$CDM baseline~\cite{Planck:2018vyg}. Since $\xi$ is strongly correlated with $S_8$, as shown in the right panel of Fig.~\ref{fig:H0-S8-xi}, this improvement can be traced back to the preference for $\xi < 0$. This trend is especially relevant given the long-standing $S_8$ tension between weak-lensing surveys and Planck assuming $\Lambda$CDM~\cite{DiValentino:2020vvd}. While the degree of alleviation remains model-dependent and a full assessment requires a direct comparison with weak-lensing measurements in the context of the IPEDE framework, the results clearly indicate that the inclusion of DESI not only reduces the Hubble constant tension but also points towards an improved agreement on $S_8$.

Next we combine three different compilations of SNe~Ia, namely PantheonPlus, Union3 and DES-Dovekie, with CMB+DESI. In all three cases, we find very strong evidence for an interaction at a high significance level, with $\xi < 0$, clearly visible in the left panel of Fig.~\ref{fig:H0-S8-xi}. This indicates that, independently of the specific supernova compilation used, the energy transfer occurs from DE to DM. The 95\% CL constraints on the interaction parameter are 
$\xi = -0.164_{-0.045}^{+0.048}$ (CMB+DESI+PantheonPlus),  
$\xi = -0.163_{-0.053}^{+0.055}$ (CMB+DESI+Union3), and  
$\xi = -0.172_{-0.041}^{+0.042}$ (CMB+DESI+DES-Dovekie). 
The mean values obtained with PantheonPlus and Union3 are very close to each other, while DES-Dovekie prefers slightly more negative values of $\xi$. This is because DES-Dovekie tends to favor a higher matter density, which in turn drives a stronger energy transfer from DE to DM (the negative correlation between $\xi$ and $\Omega_m$ can be seen in Fig.~\ref{fig:IPEDE-1}). The most conservative errors are obtained with Union3, while PantheonPlus and DESY5 show comparable uncertainties.  
The trend is also reflected in the Hubble constant: we find its 68\% CL constraints as, 
$H_0 = 68.09_{-0.59}^{+0.59}$ km/s/Mpc (CMB+DESI+PantheonPlus),  
$H_0 = 68.11_{-0.69}^{+0.69}$ km/s/Mpc (CMB+DESI+Union3), and  
$H_0 = 67.88_{-0.53}^{+0.52}$ km/s/Mpc (CMB+DESI+DES-Dovekie).  
As $\Omega_m$ increases in these combinations, $H_0$ correspondingly decreases because of the geometrical degeneracy between the two parameters. As a result, the $H_0$ tension is effectively restored.  
It is important to stress that both PEDE and the interaction can in principle alleviate the $H_0$ tension. In fact, in the CMB-only case, PEDE alone is sufficient to push $H_0$ to values in excellent agreement with local measurements. With the addition of DESI, this effect is weakened but still present. However, when supernova datasets are considered, the situation changes completely: the preference for $\xi < 0$ leads to an increase in $\Omega_m$, lowering $H_0$ and bringing the model back in tension with SH0ES.

In Fig.~\ref{fig:eff-eos-data}, we show the evolution of the effective EoS parameters with their 68\% CL credible regions for different datasets. We see that for the combined datasets, namely CMB+DESI+PantheonPlus, CMB+DESI+Union3, and CMB+DESI+DES-Dovekie, a clear transition of $w_{\rm DE}^{\rm eff}$ from the phantom to the quintessence regime is observed, i.e.\ a crossing of the phantom divide~\cite{Ozulker:2025ehg}. This behaviour is consistent with the recent DESI results~\cite{DESI:2025zgx}, which also report evidence for a dynamical dark energy component, although within the context of specific parameterizations of $w(z)$, such as the Chevallier–Polarski–Linder (CPL) form or its extensions. A growing body of analyses now supports this picture of DDE using DESI and complementary probes~\cite{Giare:2024gpk,Giare:2024oil,Wolf:2025jlc,Shajib:2025tpd,Giare:2025pzu,Kessler:2025kju,Specogna:2025guo,Cheng:2025lod,Cheng:2025hug,Lee:2025pzo}. Our results therefore provide an independent indication, within the interacting emergent DE framework, of a similar underlying trend. 
The fact that our combined datasets robustly prefer a transition of $w_{\rm DE}^{\rm eff}$ across the $-1$ boundary underscores the flexibility of interacting cosmologies in capturing features typically difficult to accommodate in standard $\Lambda$CDM extensions. 
On the other hand, except for CMB alone, all other dataset combinations consistently suggest negative values of $w_{\rm DM}^{\rm eff}$, which is again a direct consequence of the strong evidence for $\xi < 0$.

Finally, from the statistical quantities $\Delta \chi^2_{\rm min}$ and $\ln B_{ij}$ computed for all the datasets, we find that the reference models ($\Lambda$CDM and CPL) remain preferred over this emergent interacting scenario. This outcome is not unexpected, since PEDE is not a nested extension of $\Lambda$CDM or CPL, but rather forces the fit into a specific region of parameter space that is not the one preferred by the data in the standard model framework.

\begingroup
\begin{center}
\begin{table*}
\scalebox{0.85}{
	\begin{tabular}{ccccccccc}
				\hline
Parameters & CMB & CMB+DESI & CMB+DESI+PantheonPlus & CMB+DESI+Union3 & CMB+DESI+DES-Dovekie\\ \hline
				
				$\Omega_\mathrm{b} h^2$ & $0.02231_{-0.00015-0.00029}^{+0.00015+0.00030}$ & $0.02245_{-0.00014-0.00026}^{+0.00013+0.00027}$ & $0.02246_{-0.00013-0.00026}^{+0.00013+0.00027}$ & $0.02245_{-0.00013-0.00025}^{+0.00013+0.00026}$ & $0.02246_{-0.00013-0.00026}^{+0.00013+0.00027}$ \\
				
				$\Omega_\mathrm{c} h^2$ & $0.109_{-0.016-0.055}^{+0.036+0.040}$ & $0.130_{-0.0096-0.018}^{+0.013+0.018}$ & $0.135_{-0.0053-0.015}^{+0.010+0.012}$ & $0.137_{-0.0038-0.016}^{+0.010+0.011}$ & $0.1362_{-0.0047-0.015}^{+0.0099+0.012}$ \\
				
				$100\theta_\mathrm{MC}$ & $1.0413_{-0.0021-0.0025}^{+0.00093+0.0035}$ & $1.04023_{-0.00071-0.0011}^{+0.00058+0.0012}$ & $1.03996_{-0.00056-0.00088}^{+0.00044+0.0010}$ & $1.03987_{-0.00057-0.00088}^{+0.00038+0.0010}$ & $1.03992_{-0.00057-0.00086}^{+0.00038+0.00102}$ \\
				
				$\tau$ & $0.0533_{-0.0075-0.017}^{+0.0081+0.016}$ & $0.0564_{-0.0084-0.015}^{+0.0072+0.017}$ & $0.0573_{-0.0081-0.016}^{+0.0076+0.016}$ & $0.0570_{-0.0082-0.015}^{+0.0072+0.016}$ & $0.0574_{-0.0075-0.015}^{+0.0076+0.016}$ \\
				
				$n_\mathrm{s}$ & $0.9727_{-0.0044-0.0086}^{+0.0044+0.0086}$ & $0.9769_{-0.0034-0.0069}^{+0.0035+0.0072}$ & $0.9775_{-0.0036-0.0070}^{+0.0036+0.0069}$ & $0.9773_{-0.0035-0.0070}^{+0.0035+0.0070}$ & $0.9774_{-0.0034-0.0068}^{+0.0035+0.0069}$ \\
				
				$\ln(10^{10} A_\mathrm{s})$ & $3.053_{-0.016-0.033}^{+0.016+0.032}$ & $3.055_{-0.017-0.031}^{+0.015+0.034}$ & $3.056_{-0.016-0.032}^{+0.016+0.032}$ & $3.055_{-0.017-0.031}^{+0.015+0.034}$ & $3.056_{-0.016-0.031}^{+0.015+0.031}$ \\
				
				$\xi$ & $0.026_{-0.099-0.12}^{+0.063+0.14}$ & $-0.039_{-0.041-0.061}^{+0.038+0.061}$ & $-0.057_{-0.035-0.043}^{+0.019+0.050}$ & $-0.063_{-0.036-0.039}^{+0.014+0.054}$ & $-0.060_{-0.034-0.042}^{+0.017+0.050}$ \\
				
				$\alpha$ & unconst. & $>4.54$, unconst. & $>5.83>2.74$ & $>6.08>2.71$ & $>6.08>2.68$ \\
				
				$a_t$ & $0.54_{-0.20}^{+0.40}--$ & $0.456_{-0.088}^{+0.15}<0.648$ & $0.466_{-0.061-0.28}^{+0.13+0.20}$ & $0.477_{-0.059-0.25}^{+0.12+0.20}$ & $0.473_{-0.060-0.25}^{+0.13+0.19}$ \\
				
				$\Omega_\mathrm{m}$ & $0.272_{-0.072-0.15}^{+0.094+0.13}$ & $0.327_{-0.026-0.050}^{+0.031+0.048}$ & $0.344_{-0.015-0.038}^{+0.024+0.032}$ & $0.350_{-0.013-0.043}^{+0.026+0.033}$ & $0.346_{-0.013-0.038}^{+0.023+0.032}$ \\
				
				$\sigma_8$ & $0.944_{-0.26-0.29}^{+0.061+0.55}$ & $0.759_{-0.064-0.084}^{+0.038+0.097}$ & $0.730_{-0.043-0.054}^{+0.023+0.068}$ & $0.722_{-0.043-0.051}^{+0.017+0.074}$ & $0.727_{-0.039-0.054}^{+0.021+0.068}$ \\
				
				$H_0$ [Km/s/Mpc] & $70.6_{-4.8-6.4}^{+3.7+6.6}$ & $68.51_{-0.85-1.6}^{+0.75+1.6}$ & $67.89_{-0.53-1.0}^{+0.53+1.0}$ & $67.76_{-0.62-1.1}^{+0.56+1.2}$ & $67.85_{-0.53-0.94}^{+0.47+1.0}$ \\
				
				$S_8$ & $0.856_{-0.075-0.096}^{+0.023+0.16}$ & $0.789_{-0.025-0.039}^{+0.019+0.043}$ & $0.781_{-0.019-0.030}^{+0.015+0.033}$ & $0.778_{-0.019-0.029}^{+0.013+0.034}$ & $0.780_{-0.017-0.029}^{+0.013+0.032}$ \\
				
				$r_{\rm{drag}}$ [Mpc] & $147.12_{-0.30-0.58}^{+0.30+0.58}$ & $147.47_{-0.23-0.46}^{+0.23+0.45}$ & $147.51_{-0.23-0.45}^{+0.22+0.46}$ & $147.51_{-0.23-0.45}^{+0.23+0.45}$ & $147.50_{-0.23-0.42}^{+0.21+0.45}$ \\
                  \hline 
				$\Delta \chi^2_{\rm min}~({\rm w.r.t}~\text{$\Lambda$CDM})$ & 
				$1.0$ &
				$-1.1$ &
				$-4.1$ &
				$-5.1$ &
				$4.3$ \\
				$\ln B_{ij}~({\rm w.r.t}~\text{$\Lambda$CDM})$ & 
				$-0.5$ &				
				$-1.3$ & 
				$-0.6$ & 
				$0.2$ & 
				$-0.3$ \\
				\hline 
				$\Delta \chi^2_{\rm min}~{\rm (w.r.t~ CPL)}$ & 
				$-0.3$ &				
				$4.2$ &
				$1.4$ &
				$6.2$ &
				$12.9$ \\
				$\ln B_{ij}~{\rm (w.r.t~ CPL)}$ & 
				$-0.7$ &				
				$-2.2$ & 
				$-0.1$ & 
				$-3.1$ & 
				$-1.4$ \\			
                  \hline 
		\end{tabular} } 
        \caption{68\% and 95\% CL constraints on the IPEDE-gen scenario from CMB, CMB+DESI, CMB+DESI+PantheonPlus, CMB+DESI+Union3, and CMB+DESI+DES-Dovekie.}
\label{tab:iPEDE-gen} 
        \end{table*}                                  
\end{center}
\endgroup

\begin{figure*}
    \centering
    \includegraphics[width=0.85\linewidth]{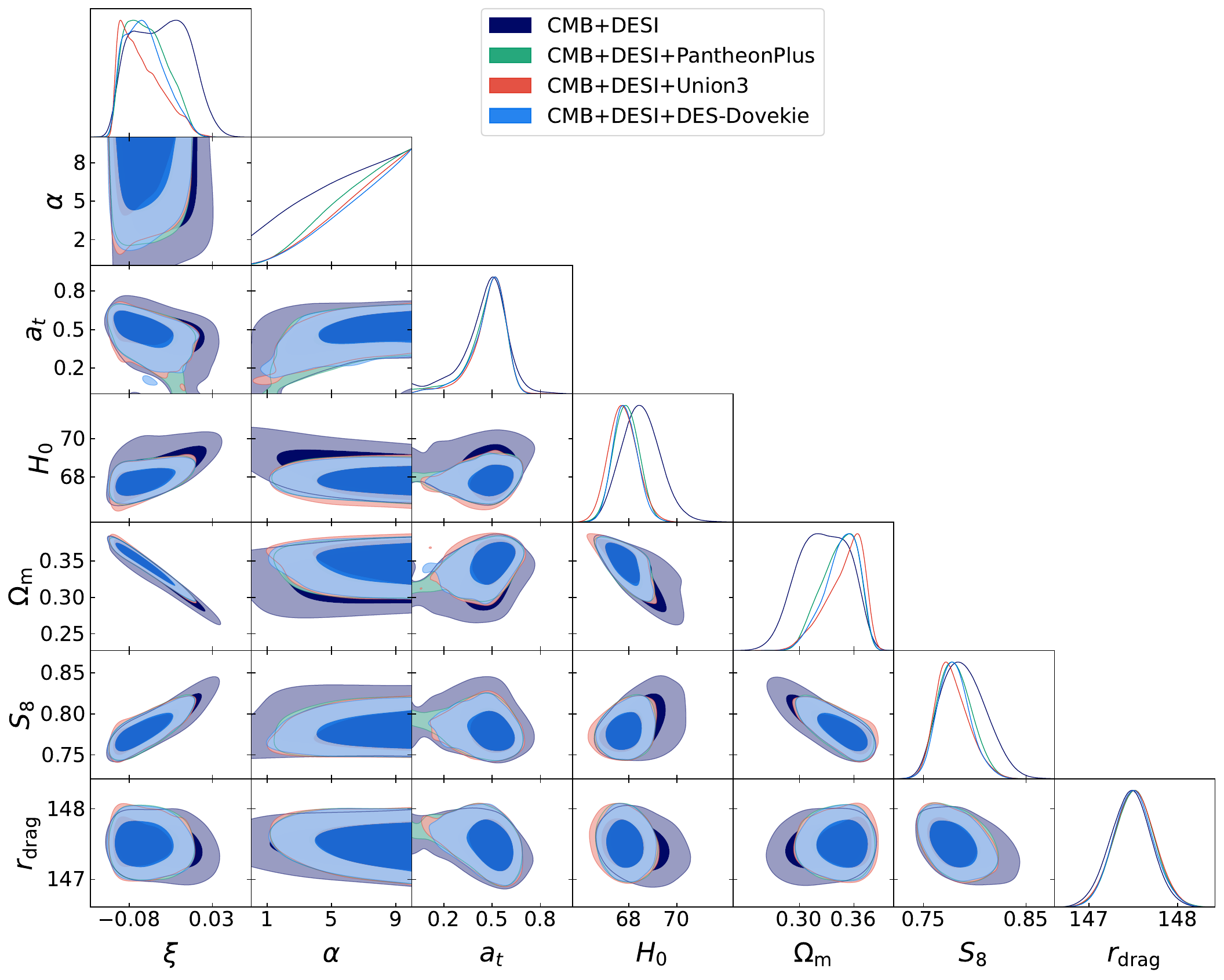}
    \caption{One-dimensional posterior distributions and two-dimensional joint contours for the most relevant parameters of the \textbf{IPEDE-gen} scenario from different combinations of cosmological datasets.}
    \label{fig:ipede-gen-alpha}
\end{figure*}

\subsection{IPEDE-gen}

In Table~\ref{tab:iPEDE-gen} and Fig.~\ref{fig:ipede-gen-alpha},\footnote{We note that in Fig.~\ref{fig:ipede-gen-alpha} we do not show the CMB-only contours, since $\alpha$ remains unconstrained in this case. Moreover, as we shall illustrate later, as $\alpha$ attains its upper limit for the remaining datasets, therefore, we do not present a plot analogous to Fig.~ \ref{fig:eff-eos-data}. } we summarize the constraints on this general interacting scenario.

Starting with the CMB-only constraints (see the second column of Table~\ref{tab:iPEDE-gen}), we find that $\alpha$ remains unconstrained, while $a_t$ is constrained only at 68\% CL ($a_t = 0.535^{+0.403}_{-0.200}$) but remains unconstrained at 95\% CL. From the constraint on $a_t$, one can mildly infer that DE turns on relatively early in this emergent interacting scenario. 
The coupling parameter $\xi$ does not show any significant evidence for a non-zero value ($\xi = 0.026^{+0.063}_{-0.099}$ at 68\% CL). However, the model leads to a high value of the Hubble constant, $H_0 = 70.6^{+3.7}_{-4.8}$ km/s/Mpc at 68\% CL, and consequently a lower value of $\Omega_m = 0.272^{+0.094}_{-0.072}$ at 68\% CL. This is mainly due to the existing anti-correlation between $\Omega_m$ and the coupling parameter (see Fig.~\ref{fig:ipede-gen-alpha}). On the other hand, the $S_8$ parameter takes a high value, $S_8 \sim 0.856$, for this dataset.

With the inclusion of DESI in combination with CMB, we find that $\alpha$ acquires a lower bound, whereas it was unconstrained in the CMB-only case. This indicates that DESI adds constraining power when combined with CMB. On the other hand, $a_t$ is constrained at 68\% CL, while at 95\% CL it attains only an upper bound. We observe a mild reduction in $a_t$ ($a_t = 0.456^{+0.15}_{-0.088}$ at 68\% CL) compared to the CMB-only case ($a_t = 0.54^{+0.40}_{-0.20}$ at 68\% CL), suggesting, similarly to CMB alone, that the emergence of DE occurs relatively early in this scenario. 
We also find a very mild indication for a non-zero coupling at slightly more than 68\% CL ($\xi = -0.039^{+0.038}_{-0.041}$ at 68\% CL). Since $\xi < 0$, the energy flow proceeds from DE to DM, leading to a higher value of the matter density parameter ($\Omega_m = 0.327^{+0.031}_{-0.026}$ at 68\% CL) and consequently a lower value of $H_0$ ($H_0 = 68.51^{+0.75}_{-0.85}$ km/s/Mpc at 68\% CL). Interestingly, in this case we also observe a lower value of $S_8$, yielding $S_8 = 0.789^{+0.019}_{-0.025}$ at 68\% CL.

For the remaining three combined datasets, namely CMB+DESI+SNe~Ia (PantheonPlus, Union3, and DES-Dovekie), we find that $\alpha$ reaches only an upper bound. From the constraints, we also note that the epoch of transition, $a_t$, lies within $\sim 0.45$--$0.47$, indicating that DE turns on relatively early in this emergent interacting scenario.
The posterior distribution of the interaction parameter is shifted away from zero at more than 95\% CL.
The corresponding 95\% CL constraints on the coupling parameter are $\xi = -0.057_{-0.035}^{+0.050}$ (CMB+DESI+PantheonPlus), $\xi = -0.063_{-0.039}^{+0.054}$ (CMB+DESI+Union3), and $\xi = -0.060_{-0.042}^{+0.050}$ (CMB+DESI+DES-Dovekie). 
 This indicates that the posterior distribution of $\xi$ remains shifted away from zero even when $\alpha$ is not well constrained. 
Since $\xi < 0$ in all cases, this indicates an energy transfer from DE to CDM, and as a consequence of this, matter density parameter takes higher values. The corresponding estimates of $H_0$ in these combined datasets are very close to the Planck $\Lambda$CDM values~\cite{Planck:2018vyg}, however, the values of $S_8$ are slightly lower ($S_8 \sim 0.778$--$0.781$) compared to the Planck $\Lambda$CDM result~\cite{Planck:2018vyg}.

We now consider the model comparison statistics. Taking $\Lambda$CDM as the reference model, we find that the $\chi^2_{\rm min}$ analysis favors {\bf IPEDE-gen} for all datasets except for CMB alone ($\Delta \chi^2_{\rm min} = 1$) and the combined dataset CMB+DESI+DES-Dovekie ($\Delta \chi^2_{\rm min} = 4.3$). However, the situation changes when CPL is used as the reference model. In this case, the $\chi^2_{\rm min}$ analysis disfavors {\bf IPEDE-gen} for all datasets, except for CMB alone ($\Delta \chi^2_{\rm min} = -0.3$)\footnote{We note that for CMB alone, $a_t$ remains unconstrained; therefore, even this very mild preference should not be considered statistically significant.}, which is nevertheless inconclusive. 
When the Bayesian evidence is taken into account, we find that the reference models ($\Lambda$CDM and CPL) are generally favored over this scenario. In particular, $\ln B_{ij} < 0$ for all datasets when CPL is used as the reference model, indicating that CPL remains preferred. On the other hand, when $\Lambda$CDM is taken as the reference model, we find $\ln B_{ij} = 0.2 > 0$ only for the CMB+DESI+Union3 combination, which corresponds to inconclusive evidence.

\section{Summary and Conclusions}
\label{sec-conclusion}

In this article we have investigated a novel interacting scenario between dark energy (DE) and cold dark matter (CDM). The novelty of this framework lies in the nature of DE, which emerges only at late times and has no effective presence in the early universe. Consequently, the interaction between DE and CDM is also absent in the past and becomes relevant only in recent epochs. This emergent behaviour of DE, together with its coupling to CDM, distinguishes the present scenario from existing works in this direction. We consider two choices for the DE equation of state, $w_{\rm DE}$: one is the well-known emergent DE model with no free parameters (Eq.~(\ref{eos}))~\cite{Li:2019yem,Pan:2019hac}, and the other is a newly proposed emergent DE model with two free parameters (see Eq.~(\ref{gen-eos})), one determining the scale factor at which DE turns on ($a_t$) and the other corresponding to the speed of the transition ($\alpha$). 
For concreteness, we have adopted a well-known interaction form proportional to the DE density, $Q = 3 H \xi \rho_{\rm DE}$, with a constant coupling parameter $\xi$.  The resulting scenarios are labeled as {\bf IPEDE} (interacting emergent DE with $w_{\rm DE}$ given in Eq.~(\ref{eos})) and {\bf IPEDE-gen} (interacting emergent DE with $w_{\rm DE}$ given in Eq.~(\ref{gen-eos})).

We have constrained both  {\bf IPEDE} and {\bf IPEDE-gen} scenarios using the latest cosmological probes: CMB, DESI BAO, and three different SNIa compilations {\bf (PantheonPlus, Union3 and DES-Dovekie)}. The observational results are summarized in Tables~\ref{tab:iPEDE} and \ref{tab:iPEDE-gen} and illustrated in Figs.~\ref{fig:IPEDE-1}, ~\ref{fig:H0-S8-xi}, \ref{fig:eff-eos-data} and \ref{fig:ipede-gen-alpha}. 
 For both interacting scenarios, posterior distribution of the coupling parameter $\xi$ is shifted away from zero only when the CMB data are combined with additional datasets. 
In particular, for all the combined datasets, the posterior distribution of $\xi$ is consistently shifted toward negative values ($\xi < 0$) corresponding to an energy transfer from DE to CDM.
However, for the {\bf IPEDE-gen} scenario, constraining the speed of the transition remains challenging with the current datasets, although $a_t$ is tightly constrained when all three datasets are combined, indicating that DE starts becoming significant at relatively early times, $a_t \sim 0.45$-$0.47$.

Regarding the cosmological parameters, CMB alone yields high values of $H_0 = 72.6^{+4.0}_{-2.8}$ km/s/Mpc at 68\% CL for IPEDE ($H_0 = 70.60^{+3.67}_{-4.80}$ km/s/Mpc at 68\% CL for {\bf IPEDE-gen}) and $S_8 = 0.86^{+0.11}_{-0.15}$ at 68\% CL for IPEDE and ($S_8 = 0.856_{-0.075}^{+0.023}$ at 68\% CL for {\bf IPEDE-gen}). This relieves the $H_0$ tension (this is effective in {\bf PEDE} compared to {\bf IPEDE-gen}) but leaves $S_8$ inconsistent with both the $\Lambda$CDM prediction from Planck~\cite{Planck:2018vyg} and recent measurements from KiDS~\cite{Wright:2025xka,Stolzner:2025htz} and DES-Y3~\cite{DES:2025xii}. However, the CMB-only constraints are not robust, being affected by known statistical degeneracies. When CMB is combined with other datasets, the inferred values of $H_0$ decrease while their uncertainties are significantly reduced. The highest value is obtained with CMB+DESI ($H_0 = 69.52 \pm 0.95$ km/s/Mpc at 68\% CL for {\bf IPEDE} and $H_0 = 68.51^{+0.75}_{-0.85}$ km/s/Mpc at 68\% CL for {\bf IPEDE-gen}), while the lowest comes from CMB+DESI+DES-Dovekie for {\bf IPEDE} ($H_0 = 67.88^{+0.52}_{-0.53}$ km/s/Mpc at 68\% CL) and from CMB+DESI+Union3 for {\bf IPEDE-gen} ($H_0 = 67.76^{+0.56}_{-0.62}$ km/s/Mpc at 68\% CL).

In contrast, the values of $S_8$ inferred from the combined datasets are significantly lower than those from Planck and recent weak-lensing surveys, with representative 68\% CL results of $S_8 \sim 0.63$--$0.69$ for {\bf IPEDE} and $S_8 \sim 0.778$ -- $0.789$ for {\bf IPEDE-gen}. This is particularly intriguing given that recent studies have suggested a consistency between Planck and lensing surveys~\cite{Wright:2025xka,Stolzner:2025htz,DES:2025xii}. Our findings therefore highlight a potential re-emergence of the $S_8$ problem in the context of interacting emergent DE.

Beyond the parameter constraints, our results highlight a distinctive behaviour of the emergent DE scenarios: with CMB alone, IPEDE successfully relieves the $H_0$ tension, with values fully consistent with local determinations (for {\bf IPEDE-gen}, although the $H_0$ tension is alleviated at slightly more than 68\% CL, this is mainly driven by the large error bars). This improvement persists, though at a reduced level, when DESI is included, but is completely lost once supernovae datasets are added. In this regard, PEDE is able to solve the Hubble constant tension when dealing with CMB data alone, still resisting with CMB+DESI, but losing in the full combination. 
 At the same time, the overall statistical comparison continues to favor $\Lambda$CDM and CPL since IPEDE and {\bf IPEDE-gen} are not nested extensions of the standard model and  hence they are penalized in the Bayesian evidence. While the posterior distributions of the coupling parameter are shifted away from zero, neither {\bf IPEDE} nor {\bf IPEDE-gen} is preferred over the reference models according to the Bayesian evidence, indicating that the improvement obtained by allowing an interaction is insufficient to compensate for the additional parameter freedom.
Although the present interacting framework is relatively complex due to the combination of an interaction model and emergent DE models, this phenomenological set-up remains appealing as it explores a specific cosmic phase in which an emergent DE component interacts with CDM and affects the late-time evolution of the universe. 
Future cosmological surveys will provide more stringent tests of these interacting emergent DE scenarios and clarify whether they can play a significant role in the late-time evolution of the Universe.

\section*{Acknowledgments}
We thank the referee for the important suggestions and comments, which have improved the manuscript.
W. Yang's work is supported by the National Natural Science Foundation of China under Grant Nos. 12547110, 12175096.  OM acknowledges the financial support from the MCIU with funding from the European Union NextGenerationEU (PRTR-C17.I01) and Generalitat Valenciana (ASFAE/2022/020). OM is also supported by the Spanish MINISTERIO DE CIENCIA E INNOVACIÓN grants PID2023-148162NB-C22 and PID2020-113644GB-I00 and by the European ITN project HIDDeN (H2020-MSCA-ITN-2019/860881-HIDDeN) and SE project ASYMMETRY (HORIZON-MSCA-2021-SE-01/101086085-ASYMMETRY) and well as by the Generalitat Valenciana grant CIPROM/2022/69.  EDV is supported by a Royal Society Dorothy Hodgkin Research Fellowship. 
This article is based upon work from the COST Action CA21136 - ``Addressing observational tensions in cosmology with systematics and fundamental physics (CosmoVerse)'', supported by COST - ``European Cooperation in Science and Technology''.


\bibliography{biblio}

\end{document}